\documentclass[10pt]{wlscirep}

\usepackage{amsmath}
\usepackage{amssymb}
\usepackage{adjustbox}

\usepackage{graphicx}
\usepackage{multirow}
\usepackage{amsmath,amssymb,amsfonts}
\usepackage{amsthm}
\usepackage{mathrsfs}

\usepackage{xcolor}
\usepackage{textcomp}
\usepackage{manyfoot}
\usepackage{booktabs}
\usepackage{algorithm}
\usepackage{algorithmicx}
\usepackage{algpseudocode}
\usepackage{listings}
\usepackage{amsmath}
\usepackage[mathlines]{lineno}
\usepackage{hyperref}

\usepackage{float}
\usepackage[utf8]{inputenc}

\usepackage{booktabs}
\usepackage{siunitx}
\usepackage{threeparttable}
\usepackage{makecell}
\usepackage{caption}
\newcommand{\zeming}[1]{\textcolor{blue}{#1}}
\newcommand{\E}{\mathbb E}

\title{Personalized optimization of pediatric HD-tDCS for dose consistency and target engagement}

\author[1,2,5,6,$\dag$]{Zeming Liu}
\author[1,3,$\dag$]{Mo Wang}
\author[1]{Xuanye Pan}
\author[5,6]{Yuan Yang}
\author[4,*]{Wilson Truccolo}
\author[1,*]{Quanying Liu}
\affil[1]{Department of Biomedical Engineering, Southern University of Science and Technology, Shenzhen, China}
\affil[2]{School of Engineering, Brown University, Providence, RI, USA}
\affil[3]{Department of Computer Science, University of Warwick, Coventry, United Kingdom}
\affil[4]{Department of Neuroscience and Carney Institute for Brain Science, Brown University, Providence, RI, USA}
\affil[5]{Carle Foundation Hospital, Urbana, IL, USA}
\affil[6]{Grainger College of Engineering, Department of Bioengineering, University of Illinois Urbana-Champaign, Urbana, IL, USA}

\affil[$\dag$]{These authors contributed equally.}
\affil[*]{Corresponding author: wilson\_truccolo@brown.edu and liuqy@sustech.edu.cn }

\begin{document}

\begin{abstract}
High-definition transcranial direct current stimulation (HD-tDCS) dosing in children remains largely empirical, relying on one-size-fits-all protocols despite rapid developmental changes in head anatomy and tissue properties that strongly modulate how transcranial currents reach the developing brain. Using 70 pediatric head models (ages 6–17) and commonly used cortical targets (primary motor cortex and left dorsolateral prefrontal cortex), our forward simulations find that standard montages produce marked age-dependent reductions in target electric-field intensity and systematic sex differences linked to tissue-volume covariation, underscoring the profound limitations of conventional uniform montages. To overcome these limitations, we introduce a developmentally informed, dual-objective optimization framework designed to generate personalized Pareto fronts summarizing the trade-off between electric-field intensity and focality. These subject-specific fronts reveal systematic performance improvements over conventional montages, yielding both higher focality at matched target intensity and higher intensity at matched focality. From these optimized solutions, we derive two clinically practical dosing prescriptions: a dose-consistency strategy that, for the first time, explicitly enforces fixed target intensity across individuals to implicitly mitigate demographic effects, and a target-engagement strategy that maximizes target intensity under safety limits. Both strategies remain robust to large conductivity variations, and we further show that dense HD-tDCS solutions admit sparse equivalents without performance loss under the target-engagement strategy. Across 1,800 optimizations in 600 conductivity-perturbed head models, we also find that tissue conductivity sensitivity is depth-dependent, with Pareto-front distributions for superficial cortical targets most influenced by gray matter, scalp, and bone conductivities, and those for a deep target predominantly shaped by gray and white matter conductivities. Together, these results establish a principled framework for pediatric HD-tDCS planning that explicitly accounts for developmental anatomy and physiological uncertainty, enabling reliable and individualized neuromodulation dosing in vulnerable pediatric populations.

\end{abstract}


\maketitle

\section{Introduction}\label{sec1}

Transcranial direct current stimulation (tDCS), as a non-invasive brain stimulation technique, uses low-amplitude direct currents to alter neuronal polarization and excitability in human brain~\cite{lefaucheur2017evidence, luo2025frequency}. High-definition tDCS (HD-tDCS), leveraging multi-electrode montages for greater spatial precision \cite{wang2023multi,dasilva2015state,lim2024high,dasilva2022concept}, has shown promise for treating neurological disorders and probing cognitive function in adults \cite{salehinejad2023optimized, ngan2022high, donnell2015high, villamar2013focal, nikolin2015focalised}. 
By contrast, pediatric studies are scarce~\cite{kessler2013dosage, andrade2014feasibility, minhas2012transcranial}: fewer than 2\% of participants in published trials are under 18 \cite{buchanan2023safety}, highlighting a major gap. Children’s brains are not scaled-down adults’: age-dependent anatomy (e.g., cerebrospinal fluid (CSF) thickness and skull composition) can systematically alter delivered electric fields~\cite{ciechanski2016transcranial, palm2016transcranial, sierawska2023transcranial}. 

Subject-specific electric-field simulation from MRI-derived head models is essential to quantify developmental and anatomical effects on dose~\cite{wang2023multi,muller2023hd,wang2025computational}. Such forward simulations can resolve between-subject field differences by explicitly parameterizing head tissues and geometry \cite{nielsen2018automatic,windhoff2013electric,huang2019realistic}. Specifically, key anatomical factors include head fat  \cite{truong2013computational}, regional CSF thickness, and skull structure  \cite{mosayebi2021impact, opitz2015determinants}, which can attenuate or deflect electric field at targeted regions. Age and sex further shape electric field strength via covarying morphology and tissue proportions~\cite{indahlastari2020modeling, antonenko2021inter}, which eventually alter stimulation outcomes \cite{rudroff2020response, bhattacharjee2022sex}. 
Therefore, to achieve maximum or consistent field strength at the target area across a diverse population, a necessary stimulation montage design is required to effectively research neuromodulation performance.

Although adult cohorts have been mined to link anatomy with fields, equivalent analyses in children remain sparse. Critically, rapid developmental change introduces substantial variability in delivered dose across children. Even when anatomical variability is acknowledged, most pediatric studies merely document rather than systematically address its influence on stimulation outcomes  \cite{dasilva2015state,wang2025hd,wang2023randomized,ma2024mapping,krauel2025prefrontal}. Only a few studies move toward individualized optimization, predominantly adult-oriented or methodological \cite{dmochowski2011optimized,ruffini2014optimization,huang2018optimized,caulfield2020transcranial,lipka2021resolving,rasmussen2021high}. Most models assume fixed tissue conductivities \cite{gabriel2009electrical, bardhi2025optimization, fabbrizzi2025reconstructing}, despite mounting evidence that inter-individual—and multi-tissue—differences reshape current flow \cite{schmidt2015impact,russell2017sex,shahid2013numerical,saturnino2019principled}. Systematic analyses of joint conductivity variability across scalp, skull, CSF, gray matter (GM) and white matter (WM) remain rare, leaving a key physiological uncertainty unresolved. However, prior analyses have primarily focused on cortical targets \cite{saturnino2019principled}, leaving open how tissue conductivity variations differentially influence current delivery to deeper brain regions, where current propagation may depend more on gray and white matter pathways. Yet, this depth-specific conductivity sensitivity has never been systematically quantified in any age group.

These anatomical and physiological factors complicate the delivery of consistent target-region fields with tDCS across individuals. To clarify the concept, we operationalize \textit{dose consistency}, the ability to achieve a prescribed target-region field across individuals under safety limits. Meanwhile, to explore the maximum achievable intensity, we define \textit{target engagement} as maximizing the feasible target-region field within safety constraints. Together, these two concepts underpin individualized HD-tDCS strategies to enhance dose consistency and target engagement in pediatric populations.

\begin{figure*}[!h]  
	\centering
	\includegraphics[width=1\textwidth]{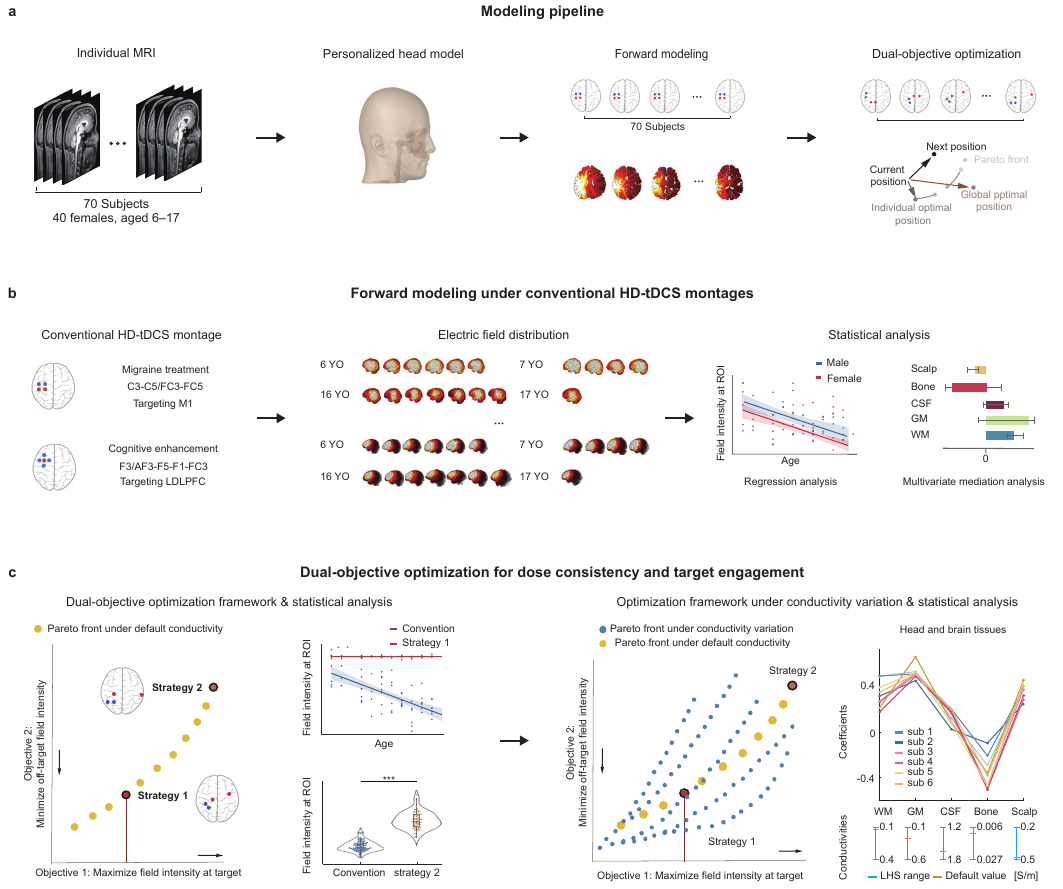}
   	\caption{Overview of the individualized HD-tDCS modeling and optimization framework.
(a) T1-weighted MRI scans (n = 70, aged 6–17) were segmented into personalized head models. These models were used for forward electric-field simulations and subsequent inverse optimization to derive subject-specific Pareto fronts under the dual-objective framework.
(b) Forward modeling with conventional HD-tDCS montages targeting M1 (C3–C5/FC3–FC5, migraine treatment) and LDLPFC (F3/AF3–F5–F1–FC3, cognitive enhancement). Electric fields were simulated for each head model, with regression and mediation analyses evaluating age/sex effects and tissue volume mediation on ROI intensity to reveal the limitations of conventional montage approaches. 
(c) Dual-objective optimization: Pareto fronts balanced maximizing ROI intensity and minimizing whole-brain mean field norm. Strategy~1 fixed ROI intensity to enhance dose
consistency; Strategy~2 maximized ROI intensity for target engagement. Statistical analyses further evaluated whether both strategies achieved their intended effects. Robustness under tissue-conductivity variation was further evaluated to assess stability and generalizability across physiological variability, along with the influence of individual tissue conductivities on Pareto fronts distributions..
	}
	\label{fig:Newfig1}
\end{figure*}

Here, we present a personalized, dual-objective optimization framework for pediatric HD-tDCS that focuses on dose
consistency and target engagement (Fig.~\ref{fig:Newfig1}a-c, Fig.~\ref{fig:S2}). We built individualized finite-element head models from 70 children and adolescents, and ran forward simulations with conventional montages to quantify age- and sex-related variability in the target fields, revealing substantial between-subject differences. We then posed an inverse optimization to generate a Pareto front that maximizes the target-region field and minimizes the whole-brain spread under safety constraints. Each point on this front represents a feasible stimulation configuration achieving a distinct trade-off between target intensity (efficacy) and focality (spread), serving as the basis for deriving individualized optimization strategies. Two downstream strategies were derived from the Pareto front: the dose-consistency strategy that selects the solution achieving a prescribed target-field level that is identical across individuals, and the target-engagement strategy that selects the solution with the maximally feasible target-field intensity on Pareto front. Relative to conventional montages, our dose-consistency strategy reduced between-subject variance and attenuated age- and sex-related trends, while the target-engagement strategy increased target fields. To assess robustness to tissue-conductivity uncertainty, we applied Latin Hypercube Sampling (LHS) to perturb scalp, skull, CSF, gray matter (GM), and white matter (WM) conductivities and re-optimized 100 times per subject in six representative children \cite{helton2003latin} targeting three target regions separately (two superficial regions and one deep brain region). Across both superficial (M1, DLPFC) and deep (hippocampus) targets, both optimization strategies remained robust under conductivity variability, with tissue sensitivity profiles exhibiting systematic shifts across target depth—from scalp- and bone-dominated effects in cortical targets to gray- and white-matter-dominated effects in deep-brain stimulation. These results support a principled route to individualized pediatric HD-tDCS planning that improves dose consistency and achievable target engagement.

\section{Results}\label{sec2}

\subsection{Age and sex significantly impact dose consistency in conventional tDCS configuration}
\begin{figure*}[!h]  
	\centering
	\includegraphics[width=0.98\textwidth]{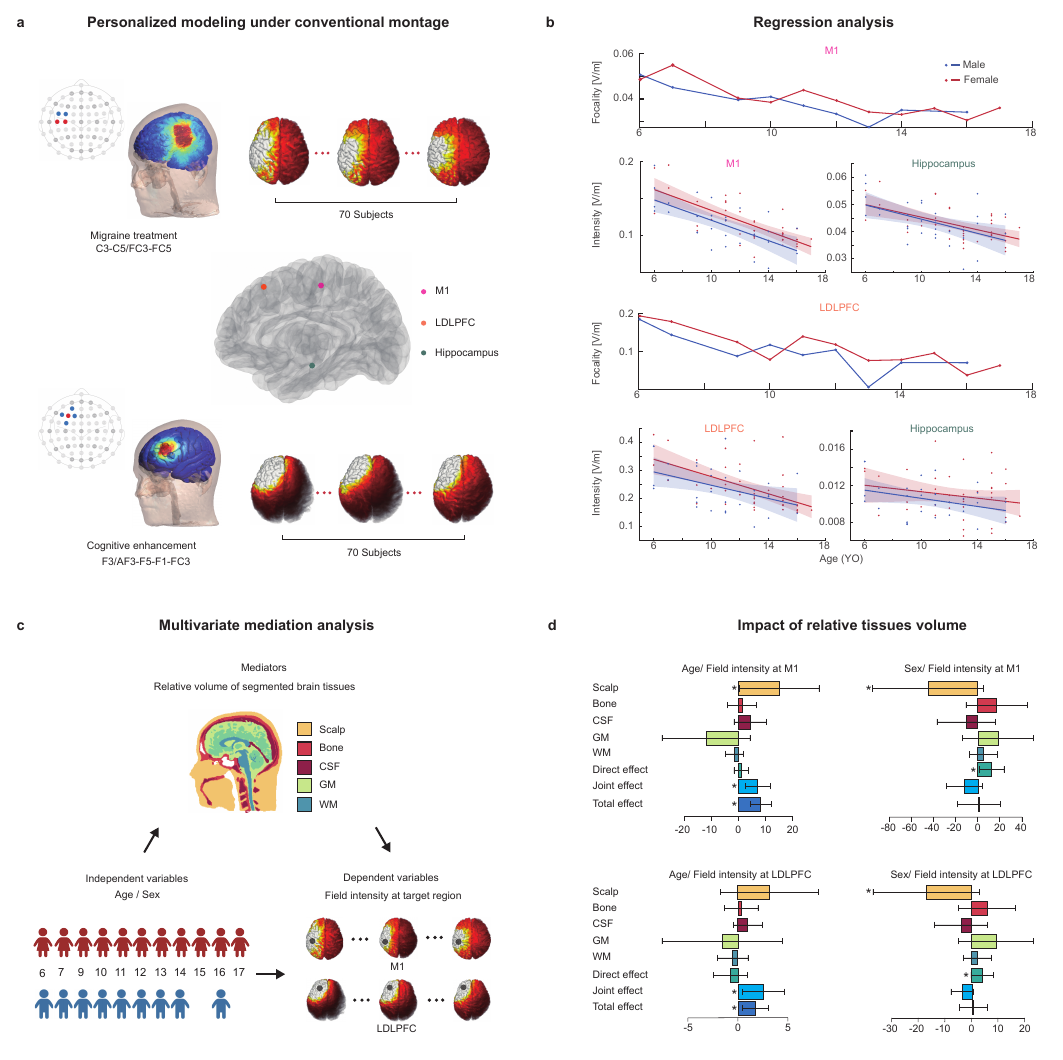}
	\caption{Electric field variability across pediatric individuals under conventional HD-tDCS montages. 
(a) Schematic of two conventional montages (M1- and LDLPFC-targeted) and corresponding field simulations across 70 pediatric head models.
(b) Regression analysis of focality (mean field norm, where lower values indicate higher focality) and ROI intensities (M1, LDLPFC, hippocampus) shows age-related decline and sex differences.
(c) Schematic of multivariate mediation analysis linking demographic variables, tissue volumes, and field intensity at targets.
(d) Multivariate mediation results decomposing age- and sex-related effects on field intensity at M1 and LDLPFC into total, direct, and tissue-specific pathways (scalp, bone, CSF, GM, WM, joint). Significant effects ($p < 0.05$) are marked with blackasterisks.
	}
	\label{fig:Newfig2}
\end{figure*}

In this study, forward modeling (Methods) was performed on 70 individualized pediatric head models (ages 6–17, 40 females) to investigate demographic and anatomical influences on electric field distributions under two conventional HD-tDCS montages:  primary motor cortex (M1)-targeted (C3–C5/FC3–FC5, migraine treatment) and left dorsolateral prefrontal cortex (LDLPFC)-targeted (F3/AF3–F5–F1–FC3, cognitive enhancement) \cite{dasilva2015state, nikolin2015focalised}. For each montage, six ROIs were defined including three superficial (LDLFPC, primary visual cortex (V1), and M1) and three deep cerebral areas (hippocampus, pallidum, and thalamus) as shown in Fig.~\ref{fig:Newfig2}a. In the main text, we focus on the montage-specific cortical target and the hippocampus as a representative deep ROI, while results for other ROIs are provided in Supplementary (Fig.~\ref{fig:S1}a--f).

Forward modeling revealed that both age and sex correlate with the induced electric field characteristics. Specifically, across both the M1 and LDLPFC stimulation montages, the whole-brain mean field and field intensity at ROIs decreased with age (Fig.~\ref{fig:Newfig2}b), indicating that younger individuals experienced more intense stimulation. Nested generalized linear models (GLMs), employing an inverse Gaussian probability density function for field intensity response, confirmed these findings, demonstrating a statistically significant age effect at all ROIs ($p<0.05$, FDR-corrected; Tables.~\ref{tab:GLM_M1_main}--\ref{tab:GLM_LDLPFC_main}) for both montages. These findings demonstrate that conventional HD-tDCS configurations produce notable inter-individual variability in current delivery at the ROIs in this pediatric population.

Given the observed demographic effects on electric field intensity, we sought to determine whether these effects were mediated by individual differences in head and brain tissue composition (Fig. ~\ref{fig:Newfig2}c). To this end, we conducted multivariate mediation analysis  \cite{yu2017mma, yu2023package} using the relative volumes of five segmented tissue types—scalp, bone, CSF, GM, and WM—as mediators (Methods). For age-related effects, scalp volume significantly mediated the reduction in electric field intensity at the M1 site. At LDLPFC, no single tissue volume was individually significant, though the joint mediation effect across all tissues reached significance, suggesting a more distributed anatomical basis for the observed age effect. Regarding sex-related effects, the anatomical pathways were similar across both stimulation strategies. Consistently, at both the M1 and LDLPFC sites, scalp volume significantly mediated the sex effect, and a significant direct effect remained (Fig.~\ref{fig:Newfig2}d). These findings underscore the critical role of scalp volume and other tissue properties in modulating electric field intensity, highlighting the importance of anatomical variability in shaping the outcomes of sex- and age-related differences under HD-tDCS stimulation.

In summary, conventional HD-tDCS montages targeting M1 and LDLPFC produce age- and sex-dependent variability in electric field distributions across both cortical and subcortical regions in pediatric populations. These findings highlight the limitations of one-size-fits-all protocols and underscore the need for individualized strategies that account for demographic and anatomical variability.

\subsection{Dual-objective optimization improves stimulation focality and intensity} \label{section2_2}

To evaluate the effectiveness of individualized optimization, we compared a conventional M1-targeted montage with subject-specific solutions derived from a dual-objective optimization procedure (Methods) that simultaneously minimized whole-brain mean field norm to improve focality and maximized M1 field intensity under fixed current and safety constraints. A personalized Pareto front was computed for each individual, with each solution representing a feasible montage reflecting a unique trade-off between the two objectives. To ensure fair comparison with the conventional montage, two representative solutions were selected per subject: one matched in M1 intensity (same intensity for comparing focality) and another matched in mean field norm across the whole brain (same focality for comparing intensity) (Fig.~\ref{fig:Newfig3}a). 

In both cases, optimized montages Pareto dominated the conventional strategy, yielding either lower whole-brain spread or stronger target stimulation (Fig.~\ref{fig:Newfig3}b-c). Mann–Whitney U tests confirmed these improvements as significant (*** $p<0.001$), indicating enhanced focality and efficacy at M1. Kernel density estimates (KDEs) illustrated inter-individual variability and sex-stratified trends in optimization outcomes (Fig.~\ref{fig:Newfig3}b-c). Relative difference was defined as the percentage change from each individual’s conventional strategy baseline. Irrespective of stimulation intensity, focality, or sex, the optimized montage consistently demonstrated superior performance. Furthermore, females exhibited a performance distribution skewed toward larger relative reductions compared to males. These results collectively establish our dual-objective optimization framework produces stimulation solutions that consistently outperform conventional fixed-montage approaches across individuals, offering a principled pathway to enhance both focality and intensity in pediatric HD-tDCS. 

\subsection{Dual-objective optimization reveals distinct strategies for dose
consistency and target engagement}

\begin{figure*}
	\centering
	\includegraphics[width=1\textwidth]{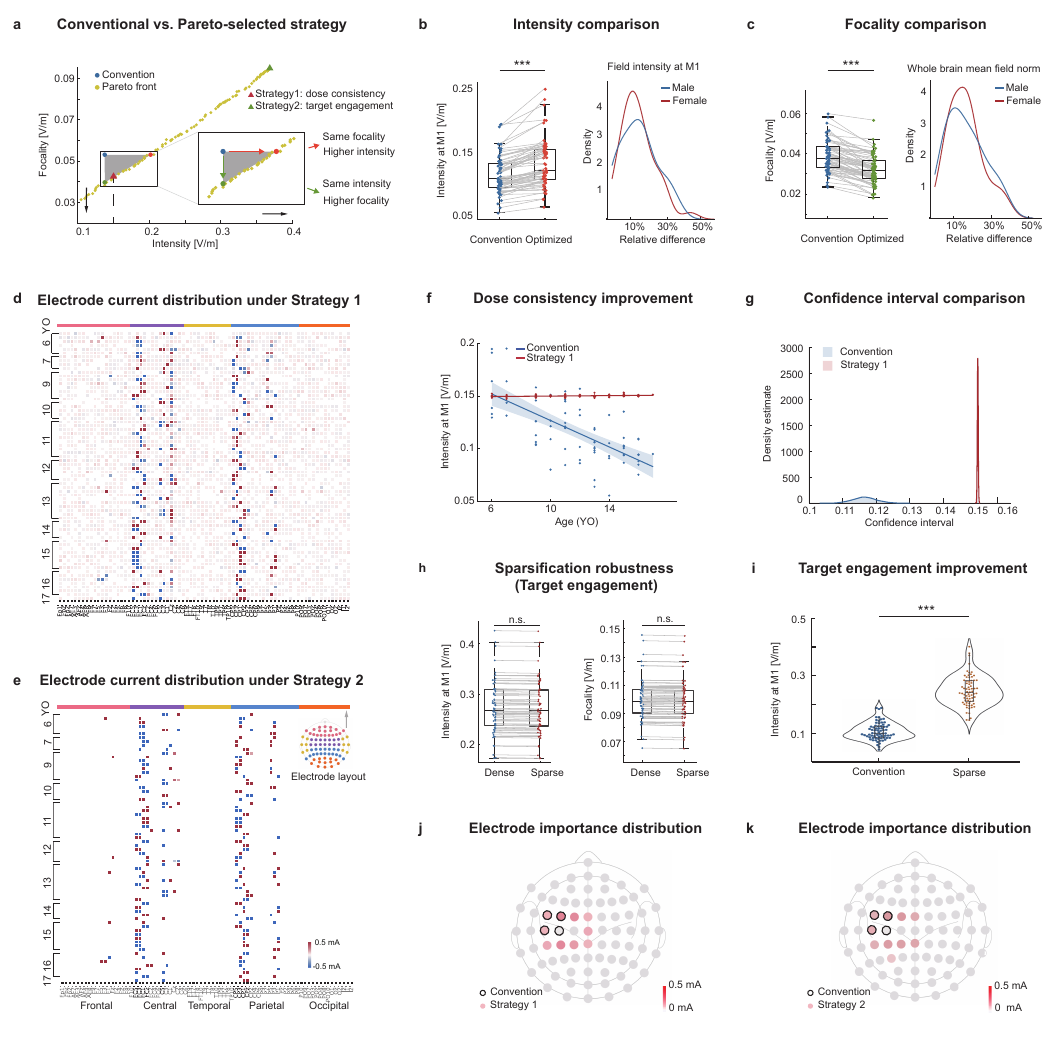}
\caption{M1-targeted optimization strategies for dose consistency and target engagement in pediatric HD-tDCS.
(a) Comparison between conventional montage and the dual-objective optimization framework. The Pareto front illustrates the trade-off between M1 intensity and focality, with blue–orange pairs highlighting intensity gains at matched focality and blue–green pairs highlighting focality improvements at matched intensity.Strategy 1 (red triangle) fixes M1 intensity at~0.15 V/m to ensure dose consistency, whereas Strategy 2 (green triangle) selects the maximal-intensity solution for target engagement. 
(b–c) Boxplots and KDEs curves further show that, compared to conventional montages, optimization significantly increased target-region (M1) intensity while also increasing focality (mean field norm, where lower values indicate higher focality) (***$p < 0.001$), with sex-stratified KDEs revealing consistent improvements across both sexes.
(d–g) Under Strategy 1, electrode current distributions reveal fixed-intensity control across individuals (d), with regression showing age-independent stability (f) and markedly narrower confidence intervals than conventional montages (g). 
(e–i) Strategy 2 produced sparse electrode configurations after 0.1 mA thresholding (e) that preserved robustness under sparsification (h) while achieving significantly higher target-region intensities than conventional montages (i, ***$p<0.001$). 
(j–k) Electrode importance analyses identified consistent fronto-central contributions under Strategy~1 and target engagement-driven optimized sites under Strategy~2.}
	\label{fig:Newfig3}
\end{figure*}

To demonstrate the optimization framework's capacity to accommodate diverse application priorities, we utilized the same procedure as described in Section \ref{section2_2} to obtain the Pareto fronts (intensity at target vs. whole-brain intensity) for both the montages targeting M1 and LDLPFC. Each participant's individualized Pareto front effectively summarizes the trade-off between intensity and focality (Fig.~\ref{fig:Newfig3}a). We subsequently defined two selection strategies from this front: Strategy 1: a dose-consistency–oriented strategy, which selects the Pareto solution corresponding to a predefined common target field value that each one could reach; and Strategy 2: a target-engagement–oriented strategy, which prioritizes the Pareto solution exhibiting the maximum target field value (Fig.~\ref{fig:Newfig3}a).

\paragraph{Strategy 1 — Dose consistency.} To evaluate whether personalized optimization improves dosing consistency, we selected for each subject the montage on their Pareto front whose target intensity was closest to a fixed reference value (0.15 V/m at M1 under 1 mA; 0.35 V/m at LDLPFC under 2 mA) (Fig.~\ref{fig:Newfig3}a, Fig.~\ref{fig:Newfig3}f, red; Fig.~\ref{fig:S3}a). This design enforces uniform stimulation intensity across individuals, thereby enhancing \textit{dose
consistency}—defined here as the ability to deliver consistent doses across anatomically diverse individuals.

The resulting montages are shown in subject-wise heatmaps (Fig.~\ref{fig:Newfig3}d; Fig.~\ref{fig:S3}b). Each row represents a subject (ordered by age) and each column a 10–10 EEG electrode, revealing individualized current patterns that consistently cluster around central–parietal electrodes for M1 and central–frontal electrodes for LDLPFC. Despite anatomical differences, this spatial convergence reflects optimization guided by head morphology while maintaining fixed target intensity.

Importantly, Strategy 1 eliminates the age and sex-related decline in field intensity at target region (M1 or LDLPFC) observed under the conventional montage (Fig. \ref{fig:Newfig2}b), underscoring its effectiveness in improving stimulation dose
consistency against age and sex effects (Fig. \ref{fig:Newfig3}f; Fig.~\ref{fig:S3}c). Nested GLMs confirmed that neither age nor sex significantly influenced intensity under this strategy (Tables.~\ref{tab:glm_personalize_M1}, top; \ref{tab:glm_personalize_LDLPFC}, top). KDEs of bootstrap-derived confidence intervals further demonstrated sharply peaked, narrow distributions compared to the broad intervals under the conventional montage (Fig.~\ref{fig:Newfig3}g; Fig.~\ref{fig:S3}d), confirming reduced variability and improved delivery precision.

\paragraph{Strategy 2 --- Target engagement.}
Complementing the dose consistency-oriented design of Strategy 1, Strategy 2 was developed to prioritize stimulation target engagement by maximizing the electric field strength specifically at the target region (Fig.~\ref{fig:Newfig3}a, Fig.~\ref{fig:S4}a). Specifically, for each subject, we selected the solution on their Pareto front that yielded the highest M1 or LDLPFC intensity under total current and safety constraints, which was subsequently sparsified by removing weakly active electrodes (\(<0.1\)~mA) (Fig.~\ref{fig:Newfig3}e and Fig.~\ref{fig:S4}b). This strategy achieved significantly higher target-region field intensity than the conventional montage (***$p<0.001$; Fig.~\ref{fig:Newfig3}i, Fig.~\ref{fig:S4}d), confirming its efficacy-oriented design. Notably, even without explicit demographic constraints, achieved target intensity showed no significant age or sex effects (Tables.~\ref{tab:glm_personalize_M1}--\ref{tab:glm_personalize_LDLPFC}, bottom), suggesting that target engagement-driven personalization can implicitly mitigate anatomical variability.
 
\paragraph{Weak-electrode pruning preserves performance while enabling clinical simplification.}

As introduced above, the optimized montages were subsequently sparsified by removing weakly active electrodes (\(<0.1\) mA), yielding individualized sparse configurations (Fig.~\ref{fig:Newfig3}e; Fig.~\ref{fig:S4}b). Importantly, this reduction did not compromise focality or intensity (n.s., Mann–Whitney U; Fig.~\ref{fig:Newfig3}h;  Fig.~\ref{fig:S4}c), while still delivering significantly higher target intensity than the conventional montage (Fig.~\ref{fig:Newfig3}i;  Fig.~\ref{fig:S4}d), despite using far fewer electrodes.

The electrode importance maps in Fig.~\ref{fig:Newfig3}j, k and Fig.~\ref{fig:S4}e compare the distribution of electrode importance between the conventional strategy and optimized strategies, targeting M1 and LDLPFC regions separately. Compared to the conventional strategy, which focuses solely on a limited set of fixed electrodes and applies the same electrode configuration across all individuals (FC3-FC5/C3-C5 or F3/AF3–F5–F1–FC3), Strategy 1 and Strategy 2 provide more focused and flexible optimized stimulation, depending on whether the goal is to stabilize the field intensity or maximize target engagement. This approach reveals that other electrodes, such as specific electrodes (e.g., FC1, FCz, CP1 for M1 or AFz, FC3, FC5 for LDLPFC), play significant roles in achieving the desired stimulation outcomes. The robustness of these optimized configurations underscores their potential for use in pediatric HD-tDCS protocols, where precise and personalized targeting is crucial for enhancing therapeutic engagements.

\subsection{Optimization strategies remain robust under variability in tissue conductivity}

\begin{figure*}[!h]  
	\centering
	\includegraphics[width=1\textwidth]{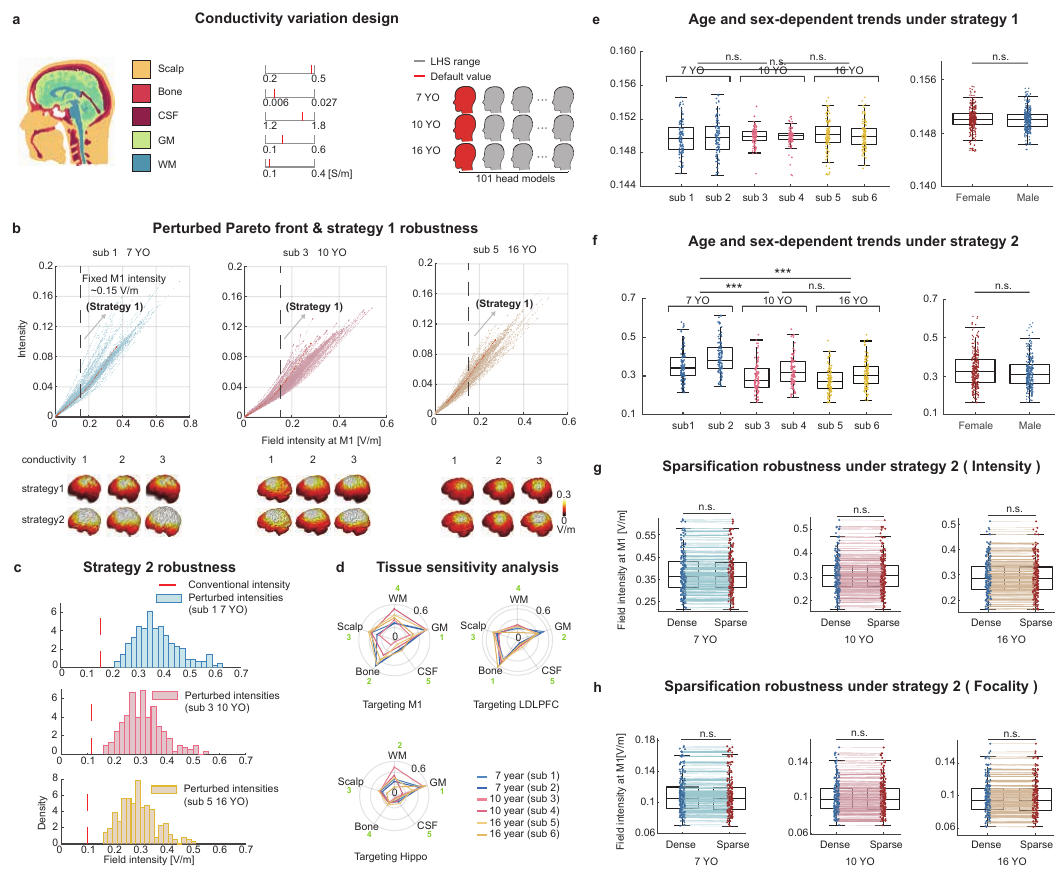}
\caption{Robustness of optimization strategies under systematic perturbations of tissue conductivity parameters.
(a) Systematic variation of conductivity parameters using a Latin Hypercube Sampling design across five tissues and three age groups.
(b) Perturbed Pareto fronts from representative individuals (ages 7, 10, and 16) illustrate the distribution of solutions under conductivity variability. For each age group, one representative individual is shown, each simulated under three conductivity samples (conductivity 1–3), with corresponding field maps under Strategy 1 and Strategy 2 below. Strategy 1 maintained a fixed target intensity (~0.15 V/m) across variations, demonstrating stable dose consistency, whereas Strategy 2 represented target engagement, achieving maximal feasible intensities under safety constraints.
(c) Strategy 2 consistently exceeded conventional montage intensities under perturbations, confirming enhanced target engagement.
(d) Tissue sensitivity analysis showed distinct conductivity dependence between cortical and deep targets: GM, scalp and bone conductivities were dominant for superficial targets (M1, LDLPFC), while gray and white matter conductivities became more influential for the deep hippocampal target, reflecting depth-dependent current propagation pathways. All coefficients represent the absolute values of OLS regression estimates, with green numbers indicating their rank order (1 = highest).
(e–f) Strategy 2 montages after sparsification preserved M1 intensity and whole-brain mean field norm, confirming robustness under large-scale variability. 
(g–h) Under variations, Strategy 1 eliminated age- and sex-related differences, whereas Strategy 2 retained residual age effects at younger ages but showed no sex differences.
}

	\label{fig:Newfig4}
\end{figure*}

Tissue conductivity will alter the distribution of electrical field across the brain. To evaluate the robustness of the dual-objective optimization framework under variability of tissue conductivity, we perturbed the conductivity parameters (Methods) for scalp, bone, CSF, GM, and WM across three representative pediatric age groups (7, 10, and 16 years) using LHS. Two individuals were selected per age group, and for each of the six individuals, 100 perturbed head models plus the default model were simulated, yielding personalized Pareto fronts that reflect realistic physiological variability (Fig.~\ref{fig:Newfig4}a). From each Pareto front, Strategy 1 and Strategy 2 were designed identically to the procedures detailed in the preceding sections. This design allowed us to directly assess how conductivity perturbations influence the stability of the two Pareto front-derived strategies.

For the M1 target, we assessed the stability of both optimization strategies (Strategy 1 and Strategy 2) under conductivity perturbations. Pareto fronts distributions from conductivity variations are shown for representative individuals across age groups (Fig.~\ref{fig:Newfig4}b). Each perturbed Pareto front (blue, pink, yellow, 100 LHS samples) is compared to the default Pareto front (red, fixed conductivities), with the vertical dashed line indicating the fixed-intensity target for Strategy 1. Across all individuals, Strategy 1 consistently achieved 0.15 V/m at M1 despite conductivity variability, demonstrating its robustness in maintaining the desired field intensity.

Fig.~\ref{fig:Newfig4}c illustrates the robustness of Strategy~2 under conductivity perturbations. For each age group (7, 10, and 16 years), histograms represent the distributions of M1 field intensity obtained from 100 perturbed models per individual (two subjects per group), with the red dashed line indicating the mean M1 intensity from the corresponding conventional montage. Wilcoxon signed-rank tests confirmed that Strategy~2 yielded significantly stronger target fields than the conventional montage across all perturbations (p $<$ 0.001). To further assess clinical feasibility, we examined the sparsified Strategy 2 montages obtained after current-thresholding under the same perturbations. As shown in Fig.~\ref{fig:Newfig4}e-f, M1 intensity and whole-brain mean field norm were statistically indistinguishable from their corresponding non-sparsified solutions (n.s., Mann–Whitney U), confirming that the Strategy 2 after current-thresholding preserves both target engagement and robustness under large-scale variability.

We next examined whether age- and sex-related effects persisted under conditions of conductivity perturbations. Importantly, demographic analyses confirmed that Strategy 1 effectively eliminated both age- and sex-related effects by enforcing fixed target intensities. In contrast, Strategy 2 preserved residual age-related differences primarily in younger participants, while demonstrating no significant sex effects (Fig.~\ref{fig:Newfig4}g–h).

We further extended the tissue sensitivity analysis beyond M1 to include the LDLPFC and the hippocampus, representing superficial and deep targets, respectively. Fig.~\ref{fig:Newfig4}d shows the Ordinary Least Squares (OLS) regression coefficients quantifying the influence of tissue conductivities on Pareto front distributions. For each individual, the Mean Ideal Distance (MID) from each front point to the ideal reference point—the maximum x-axis projection representing the strongest target intensity without current spread—was used as the dependent variable. Results revealed systematic shifts in tissue conductivity sensitivity profiles with target depth: scalp and bone dominated for cortical targets (M1, LDLPFC), whereas GM and WM contributed more strongly to deep-brain stimulation (hippocampus). These results indicate that optimization for targets at different depths depends on distinct tissue conductivity profiles, reflecting physiologically grounded differences in current propagation.

\section{Discussion}
In this study, we proposed a developmentally informed, personalized optimization framework for pediatric HD-tDCS, designed to account for demographic- and anatomy-dependent variability in the developing brain, thereby enhancing stimulation dose consistency and target engagement while ensuring robustness under tissue conductivity uncertainty. Through forward modeling of conventional uniform montages, we confirmed that electric field intensity in the target region is significantly influenced by age and sex, highlighting the limitations of one-size-fits-all dosing strategies in developmental cohorts. To address this challenge, we then applied inverse modeling—leveraging individualized head models, demographic profiling, and a dual-objective optimization strategy that simultaneously maximizes target-region intensity and minimizes off-target spread—to construct personalized Pareto fronts for each individual. These fronts enabled the principled selection of individualized solutions based on target-region control versus maximized target engagement. Specifically, we derived two downstream stimulation strategies: Strategy 1 achieves across-subject consistency in target-region dose (dose consistency) and strategy 2 maximizes stimulation intensity within each subject (target engagement). Then systematic variations in conductivity parameters further confirmed the stability of both strategies, underscoring the robustness and translational potential of our approach.

Despite their widespread use in pediatric HD-tDCS studies  \cite{dasilva2015state, kim2024exploring}, canonical montages adapted from adult protocols rely on standardized electrode placements that fail to account for the substantial anatomical variability of developing brains  \cite{ciechanski2018modeling}. Prior modeling work has shown that age and sex systematically influence field distributions  \cite{indahlastari2020modeling, bhattacharjee2022sex, ma2024mapping}, and our analyses confirmed significant age-related variability at M1 and LDLPFC (Fig.~\ref{fig:Newfig2}b; Tables.~\ref{tab:GLM_M1_main},~\ref{tab:GLM_LDLPFC_main},~\ref{tab:GLM_M1_Supplementary} and~\ref{tab:GLM_LDLPFC_Supplementary}), as well as sex and age-sex interaction effects in additional cortical and subcortical regions when targeting LDLPFC (Fig.~\ref{fig:S1}c\&f; Table.~\ref{tab:GLM_M1_Supplementary},~\ref{tab:GLM_LDLPFC_Supplementary}). These findings emphasize that uniform protocols cannot ensure reproducibility in pediatric populations.

Mediation analysis provided further insight into the anatomical basis of these effects. While scalp volume consistently mediated sex differences, the joint influence of WM, GM, CSF, bone, and scalp significantly contributed to age-related variability (Fig. \ref{fig:Newfig2}c-d). These findings expand the conventional view that CSF is the dominant determinant of age effects  \cite{opitz2015determinants,ma2024mapping,laakso2015inter,datta2009gyri,polania2018studying,antonenko2018age}, and in this pediatric population, show that scalp mediation complements prior reports attributing sex-related differences to GM volume or skull density  \cite{rampersad2011handling, koolschijn2013sex}. This inter-individual variability—driven by anatomical composition rather than montage configuration—can obscure group-level effects and limit reproducibility, underscoring the need for anatomically informed, personalized optimization strategies in pediatric HD-tDCS  \cite{ caiani2025anatomical, bikson2012computational, bardhi2025optimization}.

Notably, a recent large-scale developmental modeling study  \cite{shafiei2025reproducible} further demonstrates that cortical thickness sharply declines with age even after rigorous statistical control, whereas gray matter volume exhibits more gradual evolution. This dissociation highlights that age-related anatomical variability arises from distinct microstructural mechanisms rather than global scaling alone, supporting our finding that multiple tissue compartments jointly mediate the relationship between age and electric field intensity. Consistent with this, the mediation of age-to-field intensity by relative gray-matter volume was not significant in our models (Fig. \ref{fig:Newfig2}d).

Building on these findings, our approach generated personalized Pareto fronts for each subject by applying a dual-objective optimization framework, with each front point corresponding to a distinct stimulation montage that balances whole-brain spread and stimulation target engagement (Fig.~\ref{fig:Newfig3}a). To enable fair comparison with conventional montages, we examined representative solutions from each individual’s Pareto front. Across subjects, optimized solutions consistently outperformed the fixed montage—achieving either reduced whole-brain spread when intensity was matched, or greater target-region intensity when spread was matched (Fig.~\ref{fig:Newfig3}a-c), validating the generalizable advantage of individualized optimization-based strategies  \cite{datta2011individualized, miranda2017personalized}. Sex-stratified analyses further revealed systematic differences: females showed greater reductions in global spread, while males exhibited larger increases in M1 intensity (Fig.~\ref{fig:Newfig3}b-c). Together, these findings validate the utility of personalized dual-objective optimization as a robust framework for improving stimulation performance in pediatric HD-tDCS. Building on this foundation, to translate this framework into application-oriented strategies, we next explored two representative solutions from the Pareto fronts, each tailored to a distinct therapeutic goal.

Although prior studies have repeatedly highlighted the limitations of one-size-fits-all approaches in tDCS protocol design  \cite{evans2020dose, feng2018transcranial}, few have offered principled strategies to address these challenges  \cite{caiani2025anatomical}. In particular, studies quantitatively linking inter-individual anatomical variability to electric field distributions remain scarce  \cite{ma2024mapping} and even fewer have proposed individualized modeling frameworks to guide subject-specific stimulation protocols  \cite{ruffini2014optimization, huang2018optimized}. Building on our dual-objective optimization framework, we derived two representative strategies from each subject’s Pareto front, each tailored to a distinct design priority. The dose consistency strategy selects the solution that enforces consistent field intensity at the target region across individuals—a key requirement for experimental designs or clinical trials seeking uniform neuromodulatory exposure (Fig.~\ref{fig:Newfig3}a, f). To our knowledge, this is the first demonstration of effective dose equalization in a large pediatric cohort, yielding stimulation intensities that were statistically indistinguishable across age and sex groups, as confirmed by nested GLMs analysis (Fig.~\ref{fig:Newfig3}f; Fig.~\ref{fig:S3}c; Table.~\ref{tab:glm_personalize_M1}, top; Table.~\ref{tab:glm_personalize_LDLPFC}, top). Besides, the target engagement strategy selects the Pareto-optimal solution that maximizes target-region intensity within each individual, thereby leveraging anatomical potential while adhering to safety constraints (Fig.~\ref{fig:Newfig3}a, e, i; Fig.~\ref{fig:S4}a, b, d). This approach significantly improved stimulation strength relative to conventional strategies. Although demographic balance was not an explicit optimization goal under target engagement strategy, we found that the resulting intensities did not exhibit significant variability across age or sex groups (Table.~\ref{tab:glm_personalize_M1}, bottom; Table.~\ref{tab:glm_personalize_LDLPFC}, bottom). This suggests that anatomically informed optimization may inherently mitigate certain demographic biases  \cite{indahlastari2020modeling, bhattacharjee2022sex, laakso2015inter}.

Moreover, when a post hoc current threshold (0.1 mA) was applied to reduce the number of active channels—facilitating clinical implementation—the target engagement strategy maintained its performance, with no statistically significant differences observed in either mean field norm or target-region intensity (Fig. \ref{fig:Newfig3}h; Fig.~\ref{fig:S4}c). To our knowledge, this is the first modeling demonstration in pediatric HD-tDCS showing that sparse stimulation montages—identified via individualized optimization—can achieve target engagement comparable to dense configurations (Fig.~\ref{fig:Newfig3}h; Fig.~\ref{fig:S4}d), reinforcing the robustness of our approach and its translational potential for low-complexity, clinically deployable protocols, in line with prior theoretical work suggesting that optimal target engagement may not require dense electrode arrays  \cite{guler2016optimizing, guler2016optimization}. In addition, both strategies exhibited distinct spatial patterns in electrode importance (Fig. \ref{fig:Newfig3}j-k; Fig.~\ref{fig:S3}e; Fig.~\ref{fig:S4}e), suggesting that effective ROI-targeted HD-tDCS in pediatric populations may benefit from stimulation configurations beyond conventional adult-derived montages. These findings emphasize the importance of goal-specific, anatomically guided montage design in pediatric HD-tDCS—moving beyond one-size-fits-all templates to develop stimulation strategies that are both effective and clinically feasible. While our framework was demonstrated in pediatric populations, the underlying principles of anatomically-informed dual-objective optimization and montage sparsification is readily extensible to other age groups or clinical populations, paving the way for broader applications of individualized neuromodulation strategies across diverse populations.

Furthermore, we demonstrated that both Strategy 1 (dose consistency) and Strategy 2 (target engagement) remained robust under conductivity perturbations (Fig.~\ref{fig:Newfig4}a-c). Strategy 1 maintained consistent dosing across individuals, whereas Strategy 2 maximized target intensity and further validated by the ability to reduce the number of active channels without compromising performance (Fig.~\ref{fig:Newfig4}e-f). Demographic analyses showed that Strategy 1 abolished age- and sex-related effects, as expected from its fixed-intensity design, whereas Strategy 2, while targeting target engagement, also attenuated age-related variability and eliminated sex effects (Fig. \ref{fig:Newfig4}g-h). This demonstrates that our framework not only ensures dosing uniformity when desired but can also reduce part of the demographic influences even under target engagement-driven optimization. 

Sensitivity analyses further revealed a depth-dependent hierarchy in tissue conductivity influence: scalp and bone conductivities dominated for cortical targets (M1, LDLPFC), whereas gray and white matter conductivities exerted stronger effects for the deep target (hippocampus) (Fig.~\ref{fig:Newfig4}d). These findings align with prior forward-modeling studies showing that GM, bone, and skin conductivities are dominant contributors to field variability when  targeting M1, as quantified using non-intrusive generalized polynomial chaos expansion \cite{saturnino2019principled}. Thus, our results show that the inverse optimization framework can capture the same biophysical sensitivities. However, to the best of our knowledge, no previous study has systematically examined whether these tissue-conductivity dependencies remain consistent—or shift—when targeting deeper brain structures. Our results reveal a clear transition in the relative importance of tissue conductivities with target depth, suggesting that optimization for superficial versus deep stimulation engages distinct current propagation pathways shaped by different tissue compartments. Beyond migraine and cognitive neuromodulation examined here, we envision the proposed framework to easily extend to many other therapeutic interventions in neurologic and psychiatric disorders, including the prevention of epileptic seizures and stroke\cite{islam2025noninvasive,moosavi2025controllability,liang2022online, williamson2025neuroengineering, williamson2023high, peng2024determining}, for example. Together, our findings highlight the capacity of the proposed framework to adapt to real-world clinical scenarios, where tissue conductivities may vary substantially across individuals, paving the way for more personalized and effective neuromodulation treatments.

In summary, this study used forward modeling to reveal age- and sex-related variations in electric field distribution during HD-tDCS in a pediatric population and explored the mediating role of head and brain anatomical structures. The findings demonstrate that age significantly influences electric field intensity, with this effect partially mediated by head and brain anatomical structures, while sex-related variations are region- and strategy-specific, further emphasizing the critical role of developmental and anatomical factors. These results underscore that adult HD-tES strategies cannot be directly applied to pediatric populations, highlighting the necessity of developing personalized stimulation protocols tailored to the developing brain. To address this, we employed the Multi-Objective Evolutionary Algorithm (MOVEA) to design individualized stimulation parameters, achieving significant improvements in target region intensity and overall stimulation target engagement while mitigating variability caused by age and sex differences. Additionally, systematic variation of conductivity parameters identified GM, scalp, and bone conductivities as significant factors influencing the distribution of Pareto fronts, further emphasizing the critical role of anatomical and biophysical properties in tES optimization. These results provide a scientific foundation for the precise application of HD-tES in pediatric populations and lay the groundwork for developing more effective personalized stimulation strategies.

\section{Methods}
\label{Methods}
This study integrates subject-specific forward modeling and dual-objective inverse optimization to design individualized HD-tDCS strategies for a pediatric population. Prior studies have demonstrated the target engagement of fixed montages—C3–C5/FC3–FC5 for M1 targeting in migraine treatment  \cite{dasilva2015state}, and F3/AF3–F5–F1–FC3 for targeting the LDLPFC to enhance declarative language learning and memory function  \cite{nikolin2015focalised}. Building on these designs, we evaluated how demographic and anatomical variables—age, sex, and tissue volumes—influence electric field strength across both superficial and deep brain regions.

To overcome the limitations of one-size-fits-all designs, we applied a dual-objective inverse optimization framework to derive personalized stimulation solutions that balance target-region intensity and global focality under safety constraints. The framework yields two distinct strategies tailored to different design priorities: Strategy 1 enforces inter-subject consistency (dose consistency), whereas Strategy 2 maximizes within-subject target engagement.

To examine the robustness of these strategies, we systematically perturbed conductivity values across five tissue types using LHS. We then assessed whether optimization goals remained attainable and analyzed how tissue-specific variability affected inter-subject differences in the resulting Pareto fronts.

\subsection{Computational Modeling}

We implemented the proposed framework using finite element method (FEM) based on individualized MRI-derived head models, with electrode placements aligned to the international 10–10 system.

For each subject, we performed the following MRI-based preprocessing steps: (I) electrode alignment to the target space and ROI definition; (II) automated tissue segmentation and extraction of corresponding volume parameters; After preprocessing, we (III) conduct the HD-tDCS forward simulation through FEM models. The above steps are based on the SimNIBS software (version 3.2.6)  \cite{saturnino2019simnibs,thielscher2015field,windhoff2013electric}. (IV) Then we use FEM models and ROIs as input to run the dual-objective inverse optimization algorithm and also conduct conductivity parameter assessment.

\subsubsection{Forward Modeling}

\paragraph{Alignment of electrodes with the target space and ROI definition}
In conducting electroencephalography (EEG) studies and high-definition transcranial electrical stimulation (HD-tES) experiment designs, the precise localization of electrodes is paramount. This study uses four reference points (LPA, RPA, Nz, Iz) to pinpoint EEG coordinates within the 10-10 EEG positions  \cite{jurcak200710}, ensuring electrode precision and alignment with brain regions and HD-tES sensors.

In the selection of ROIs, we focused on three primary regions of interest—two superficial areas, the M1 and the LDLPFC, and one deep structure, the hippocampus. These regions were selected based on their clinical relevance in pediatric neuromodulation applications and were defined according to the Brodmann Atlas  \cite{lacadie2008brodmann}. Additional analyses of other superficial and deep regions, including the primary visual cortex (V1), thalamus, and globus pallidus, are provided in Supplementary Materials (Fig. \ref{fig:S1}).

\paragraph{Tissue volume estimation}
The Headreco toolbox, integrating the functionalities of both the SPM12 and CAT12 toolboxes, facilitates the automated segmentation of various tissue types (WM, GM, CSF, bone, and scalp)  \cite{nielsen2018automatic,windhoff2013electric}. Each dataset underwent visual examination to confirm the precision of head reconstructions and tissue segmentation as recommended  \cite{saturnino2019simnibs}. Following this review, all head models were considered suitable, eliminating the need for any manual adjustments to the head reconstructions.

To characterize anatomical variability across subjects, we calculated the relative volume of each tissue as its absolute volume divided by the total head volume. These normalized values were used in subsequent statistical and mediation analyses to assess their contributions to electric field variability.

\paragraph{Electric field simulation}
Numerous techniques have been formulated for creating an authentic head model, among which are the boundary element method and the FEM  \cite{wang2025computational,laakso2012fast}.  We opted to utilize the FEM technique owing to its heightened precision and computational effectiveness in this investigation  \cite{saturnino2019electric}, facilitating the application of the Laplace equation to simulate electric potential in the brain's conductive medium  \cite{neuling2012finite}. This approach enables precise modeling of the brain's complex geometry and electrical properties, critical for studying the effects of extrinsic electrical stimulation. Given the brain's electrical signals are too weak to be considered in this context, we simplified the model to one without sources or sinks, allowing for a more focused study on the effects of external electrical fields  \cite{dasilva2015state,griffiths2018college,miranda2006modeling,wagner2007transcranial}, as Eq. (1) shows
\begin{equation}
\label{eq:Laplace}
-\nabla \cdot \left( {\sigma \nabla V} \right) = 0,
\end{equation}
where $V$ represents the electric potential field generated within the brain's conductive medium under external electrical stimulation, $\sigma$  is the isotropic electrical conductivity ($S/m$) assigned to different head and brain tissue and spaces as: scalp: 0.465, bone: 0.01, CSF: 1.65, GM: 0.276, WM: 0.126, saline: 1.4, air cavities: 2.5 $e^{-14}$, eye: 0.5, according to previous literature \cite{dasilva2015state,opitz2015determinants,wagner2004three,saturnino2015importance}.

Following the international 10/10 electrode placement system, 76 HD-tES electrodes were strategically positioned on the scalp of the FEM model, with a reference electrode located at Cz, using SimNIBS 3.2.6  \cite{saturnino2019simnibs}. Circular saline electrodes with diameters of 12 mm and 20 mm were applied respectively for M1 stimulation in migraine treatment  \cite{dasilva2015state} and LDLPFC stimulation for declarative language learning and memory functioning  \cite{nikolin2015focalised}, both with a thickness of 5 mm. In the forward model, M1 stimulation used anodes at C3 and C5 and cathodes at FC3 and FC5 with 1 mA total current, while LDLPFC stimulation positioned the anode at F3 and cathodes at AF3, F5, FC1, and FC3 with 2 mA total.

The computation of the electric field induced by tDCS involves determining the field generated within the brain by the input current \( s \) applied through the electrodes, excluding the reference electrode. The electric field can be expressed using the lead field matrix \( A \) as \( E = As \), where \( E \) represents the triaxial electric field intensity vector. The lead field matrix encapsulates the relationship between the input current applied through HD-tES electrodes and the resulting electric field distribution across brain regions. It provides a mapping that predicts the electric field generated in each voxel of the brain for a unit input current at the electrode level. Following the method of  \cite{saturnino2019simnibs}, the average electric field intensity across a target region \( \Omega \), denoted as \( E_\Omega \), is defined as:

\begin{equation}
\begin{aligned}
E_\Omega &= \frac{1}{G_\Omega} \int_\Omega \|E\| \, dG = \frac{1}{G_\Omega} \int_\Omega \sqrt{w_x E_x^2 + w_y E_y^2 + w_z E_z^2} \, dG,
\end{aligned}
\end{equation}
where \( G_\Omega \) represents the total volume within the region \( \Omega \) in the brain. The variables \( x \), \( y \), and \( z \) denote three orthogonal directions, while \( w_x \), \( w_y \), and \( w_z \) are weights assigned to each respective direction. In this study,  the weights \( w_x \), \( w_y \), and \( w_z \) are set to unity.

\subsubsection{Dual-objective Optimization to search the optimal stimulation solutions}

To move beyond the limitations of conventional, fixed electrode montages, we formulated the electrode configuration problem as a constrained dual-objective optimization task. This framework simultaneously maximizes electric field intensity at the target region while minimizing its spatial spread across the whole brain, under strict safety constraints on injected current. The trade-off between focality and intensity is particularly critical in pediatric populations, where anatomical variability demands individualized stimulation solutions.

We adapted a previously developed optimization framework  \cite{wang2023multi} by constraining the objectives to two competing goals: maximizing target-region intensity and minimizing global field spread, and applied it for the first time to the optimization of HD-tDCS configurations. The lead field matrix derived from each subject’s forward model serves as input. The optimization process consists of two stages: an initial single-objective search using a genetic algorithm to identify a high-intensity solution at the target, followed by a multi-objective particle swarm optimization (MOPSO) stage that generates a subject-specific Pareto front of optimal solutions  \cite{yang2009novel}, reflecting the set of trade-off solutions—none of which can improve one objective without compromising the other. Each point on the front represents an individualized electrode configuration that achieves a unique balance between intensity and focality. From each front, we selected two representative solutions corresponding to distinct design priorities. Strategy 1 enforces inter-subject consistency by selecting a common fixed field intensity across individuals (dose consistency), while Strategy 2 maximizes individual target intensity (target engagement). These strategies allow us to investigate how optimization goals shape montage configurations and to compare their robustness under individual anatomical variability.

We implemented the dual-objective optimization separately for two clinically relevant targets: the M1 and the LDLPFC. Candidate electrode placements were constrained to the international 10–10 EEG coordinate system to ensure compatibility with standard HD-tES caps. To maintain consistency with forward-modeling setups and prior clinical protocols, the total injected current was set to 1 mA for M1 targeting and 2 mA for LDLPFC targeting  \cite{dasilva2015state, nikolin2015focalised}. Safety constraints were enforced such that the current delivered to any single electrode did not exceed 0.5 mA for M1 stimulation and 1 mA for LDLPFC stimulation.

Beyond anatomical variability, the biophysical properties of head tissues—particularly their electrical conductivities—also influence stimulation outcomes. Prior studies have demonstrated that conductivity values vary substantially across individuals and tissues, with further modulation by demographic factors such as age and sex \cite{schmidt2015impact, mccann2019variation}. This is especially critical in pediatric populations, where such variations may significantly affect electric field distributions. To assess the robustness of our dual-objective optimization framework under these uncertainties, we introduced systematic perturbations to tissue conductivity values using LHS, a method widely used for efficient exploration of multidimensional parameter spaces \cite{marino2008methodology}. By simulating across a uniformly sampled range of physiologically plausible conductivities (detailed below), we evaluate how variability in tissue properties influences the distribution of individual Pareto fronts and, more importantly, whether representative stimulation strategies—Strategy 1 (dose consistency) and Strategy 2 (target engagement)—can still be reliably identified across perturbed conductivity conditions. This allows us to assess the robustness of strategy selection under physiological uncertainty.

\subsection{Statistical analysis}
The overarching aim of the present study separates into two parts. First, in the forward-model simulation experiments, we investigate the effect of age and sex (both main and interaction effects) on electric field intensity at specified ROIs and the underlying anatomical factors mediating it. Second, in the inverse-model simulation experiments, we start by evaluating our optimized stimulation strategies by comparing focality and intensity enhancements over the conventional strategy. Then, we select a specific configuration on the Pareto front to address age-related limitations identified in the forward model for each participant. Finally, we quantify the uncertainty in Pareto front distributions arising from variations in the conductivity parameters for different head and brain tissue types. 

Hence, several statistical analyses were performed. In the forward experiments: (1) Assuming that age and biologic sex impact ROIs electric field strength and interact with each other, nested generalized linear models were created to evaluate the individual and combined interaction effects on electric field strength of these two factors. (2) To investigate whether the relationship between age, sex, and electric field strength at ROIs is mediated by cortical anatomical factors, we performed multivariate mediation analyses with multiple mediating variables, examining the anatomical pathways through which age and sex may exert their effects on electric field strength. In the inverse experiments: (3) To assess the statistical significance of focality and intensity improvements under the optimized versus conventional strategies, we conducted Mann-Whitney U tests. (4) For each individual, we selected both Strategy 1 and Strategy 2 from the optimized Pareto front and applied nested GLMs to determine whether the optimized approaches mitigated the significant age-, sex-, and age–sex interaction effects identified in the forward-model analysis. (5) Finally, using the LHS method, we generated 100 sets of conductivity parameters within specified ranges for each tissue type. We first examined whether Strategy 1 and Strategy 2 remained robust under conductivity variations. We then quantified the distribution of the optimized Pareto fronts across these parameter sets by employing the Mean Ideal Distance (MID) metric. Ordinary linear regression was then used to assess the influence of tissue conductivity on Pareto front outcomes, with model coefficients indicating the relative contribution of the conductivity of each tissue type to the variability in the optimization results. The above steps are described in more detail below.

\subsubsection{Forward modeling statistical analysis}
\paragraph{Modeling age and sex effects on electric field intensity under conventional montage stimulation}

Nested GLMs were estimated with the glmfit function in Matlab (MathWorks, Inc.), focusing on both the main and interaction effects of age and sex on the electric field intensity at specified ROIs. In this analysis, sex is encoded as a dummy variable, ensuring its categorical nature is appropriately handled in the model. The response variable (electric field intensity) in the nested GLM models was assumed to be distributed according to the inverse Gaussian distribution, with the corresponding  link function given by \( \frac{1}{\mu^2} \). This distribution is appropriate for modeling continuous positive response variables \cite{dunn2018chapter}. By comparing the deviance differences between nested models and calculating the corresponding \textit{p}-values from the chi-square test, adjusted for multiple testing via the False Discovery Rate (FDR) correction  \cite{benjamini1995controlling}, we can determine whether the added variables have a significant impact on the electric field intensity at the ROIs.

The nested generalized linear models were formulated as:
\[
\begin{aligned}
M_0: &\quad \E[Y] = f(\beta_0) \\
M_1: &\quad \E[Y\vert X] = f(\beta_0 + \beta_1 \times \text{Age}) \\
M_2: &\quad \E[Y\vert X] = f(\beta_0 + \beta_1 \times \text{Age} + \beta_2 \times \text{Sex}) \\
M_3: &\quad \E[Y\vert X] = f(\beta_0 + \beta_1 \times \text{Age} + \beta_2 \times \text{Sex} 
        + \beta_3 \times (\text{Age} \times \text{Sex})) \\
M_4: &\quad \E[Y\vert X] = f(\beta_0 + \beta_1 \times \text{Age} + \beta_2 \times \text{Sex}
        + \beta_3 \times (\text{Age} \times \text{Sex})
        + \beta_4 \times (\text{Age} \times (1-\text{Sex}))).
\end{aligned}
\]

where: 
\begin{itemize}
    \item $\E[Y]$ is the unconditional expected value of the electric field intensity $Y$, and $\E[Y \vert X]$ is the expected value of the field intensity conditioned on a set of predictors $X$ which is different for each different model. The electric field intensity is at a given ROIs under a given HD-tDCS stimulation strategy. The R0Is included three superficial regions (LDLFPC, V1, and pallidum) and three deeper regions (M1, hippocampus, and thalamus).
    \item $f(\cdot)$ denotes the inverse link function associated with the inverse Gaussian distribution, i.e., \( f = g^{-1} \), where the canonical link function is \( g(\mu) = \frac{1}{\mu^2} \).
    \item $\beta_0$ is the intercept, representing the baseline of the stimulation-driven electric field intensity.
    \item $\beta_1$ is the coefficient for age, quantifying the change in electric field intensity per year of age.
    \item $\beta_2$ is the coefficient for the sex dummy variable, which is encoded as 0 for females and 1 for males, indicating the difference in electric field intensity between females and males.
    \item $\beta_3$ is the coefficient for the interaction between age and sex (specifically age and males, since $\text{Sex} = 1$ for males). It shows how the effect of age on electric field intensity differs between males andfemales.
    \item$\beta_4$ further refines the interaction, indicating the effect of age when sex is coded as female ($\text{Sex} = 0$). In this case, $\beta_4$ captures how age affects intensity differently in females compared to males.

\end{itemize}

\paragraph{Evaluating the mediation of anatomical factors to sex and age effects on across-subjects HD-tDCS variability}

We used multivariate mediation analysis  \cite{baron1986moderator}, implemented in the ''mma'' package in the R software  \cite{yu2017mma, yu2023package}, to determine whether the relationship between age and sex factors the and the electric field intensity was mediated by the volumes of the following tissues WM, GM, CSF, skull (bone) or scalp. Accordingly, age and sex were used as independent variables, target region intensity as the dependent variable, and the rest of the volume parameters as well as their joint effect as potential mediators. Mediation analysis incorporating multiple mediators enables the evaluation of the distinct effects of each parameter as well as the combined effects of multiple parameters. As stated above for the estimation of nested GLMs, in this analysis the dependent variables were manually specified to follow an inverse Gaussian distribution, with a link function given by \( \frac{1}{\mu^2} \), as this distribution was deemed more appropriate for accurately capturing the intensity values observed in the data  \cite{dunn2018chapter}. To assess the significance of the mediators, the bootstrap method with 5,000 resampling iterations was employed, following the recommendations of Preacher et al \cite{preacher2008asymptotic}.

\begin{table*}[!h]
\centering
\small
\caption{Ranges for electrical conductivity parameters assigned to the different tissue types (in S/m). WM: white matter; GM: gray matter; CSF: Cerebrospinal fluid.}
\begin{threeparttable}
\begin{tabular}{l
                S[table-format=1.3]
                S[table-format=1.3]
                >{\raggedright\arraybackslash}p{12.3cm}}
\toprule
\textbf{Tissue} & \textbf{Min} & \textbf{Max} & \textbf{Reference(s)} \\
\midrule
WM   & 0.100 & 0.400 & \makecell[tl]{Jia et al. (2022)\, \cite{jia2022deeptdcs};\ Akhtari et al. (2010)\, \cite{akhtari2010variable}} \\
GM   & 0.100 & 0.600 & \makecell[tl]{Li et al. (1968)\, \cite{li1968specific};\ Saturnino et al. (2019)\, \cite{saturnino2019principled}} \\
CSF  & 1.200 & 1.800 & \makecell[tl]{Saturnino et al. (2019)\, \cite{saturnino2019principled};\ Baumann et al. (1997)\, \cite{baumann1997electrical}} \\
Skull& 0.006 & 0.027 & \makecell[tl]{McCann et al. (2019)\, \cite{mccann2019variation};\ Oostendorp et al. (2000)\, \cite{oostendorp2000conductivity};\ Fernández et al. (2016)\, \cite{fernandez2016effects}} \\
Scalp& 0.200 & 0.500 & \makecell[tl]{Fernández et al. (2016)\, \cite{fernandez2016effects};\ Fernández et al. (2017)\, \cite{fernandez2017skull};\ Bashar et al. (2010)\, \cite{bashar2010uncertainty}} \\
\bottomrule
\end{tabular}
\end{threeparttable}
\end{table*}

\subsubsection{Dual-objective inverse optimization statistical analysis}
\paragraph{Comparative analysis of optimized and conventional HD-tDCS strategies}

We assessed the target engagement of optimized stimulation strategies generated via the dual-objective inverse optimization framework by comparing electric field intensities targeting M1 with those from the conventional 2x2 tDCS montage (C3-C5/FC3-FC5). For each participant’s Pareto front, we selected two points for comparison: one point with a mean electric field norm closest to that of the conventional montage, used to compare M1 intensities, and another point with an M1 intensity closest to that of the conventional montage, used to compare mean field norms. We note that the lower the mean field norm, the highest the focality of the HD-tDCS. We then conducted Mann-Whitney U tests to identify potential improvements in focality and intensity provided by the personalized strategies. 

To ensure a valid comparison, this analysis was restricted to the M1-targeted condition, where both the conventional and optimized strategies shared the same total injected current (1 mA) and complied with per-electrode safety limits ($\leq$ 0.5 mA). For LDLPFC targeting, the conventional 4$\times$1 montage concentrated the full 2 mA at a single anode, exceeding the safety constraints enforced in the optimization framework.

\paragraph{Verifying the intended effects of dose consistency and target engagement strategies}

To confirm that Strategy~1 (dose consistency) and Strategy~2 (target engagement) successfully achieved their respective design goals, we examined electric field intensity distributions across individuals. Both strategies were selected from each individual's subject-specific Pareto front: Strategy~1 corresponds to a solution enforcing a fixed electric field intensity across individuals (\zeming{0.15 \& 0.35 $V/m$}), while Strategy~2 selects the point at the upper-right corner of the Pareto front, representing the configuration that maximizes target-region intensity within safety limits. For Strategy~2, we further excluded weakly activated electrodes by applying a 0.1~mA threshold, ensuring focus on the most effective channels and enhancing potential clinical interpretability and translational relevance of the resulting montages.

For Strategy~1, we tested whether electric field intensity at the target region was statistically indistinguishable across age and sex groups, indicating enforced uniformity. This was assessed using nested GLMs with an inverse Gaussian distribution and a link function of \( \frac{1}{\mu^2} \), where age and sex were included as predictors. For Strategy~2, we first assessed the feasibility of applying a 0.1~mA threshold to exclude weakly activated electrodes. Specifically, we conducted Mann–Whitney~U tests to evaluate whether this thresholding procedure had a significant impact on the mean field norm and the target-region intensity, thereby ensuring the stability of key stimulation metrics. Upon confirming feasibility, we then evaluated whether the field intensity at target-region under Strategy~2 was significantly higher than that observed under the conventional montage using Mann–Whitney~U tests, reflecting the strategy’s goal of maximizing stimulation target engagement. In addition, we applied the same nested GLM framework under Strategy~2 to investigate whether the intensity-maximizing configuration indirectly attenuated or eliminated the effects of age and sex on electric field outcomes.

Finally, to assess the relative contribution of each electrode under different strategies, we computed the mean absolute injected current at each electrode across all 70 individuals for both Strategy~1 and Strategy~2, allowing visualization of electrode importance under each design priority.

\paragraph{Quantitative impact of the uncertainty in the choice of head and brain tissue conductivity parameters on Pareto front distributions}

Inter-individual variability in head and brain tissue conductivities introduces uncertainty into electric field modeling and subsequent optimization outcomes. Previous studies have documented significant variability in conductivity values across different tissues, particularly in pediatric populations where developmental changes are ongoing  \cite{schmidt2015impact, mccann2019variation}. To assess the robustness of stimulation strategies derived from the dual-objective optimization framework under such uncertainty, we systematically perturbed the conductivities of five tissue types (WM, GM, CSF, bone, and scalp) using Latin Hypercube Sampling (LHS), uniformly sampling across physiologically plausible intervals (Table 1).

We selected six representative individuals spanning three developmental stages (ages 7, 10, and 16 years), with two individuals per group. These ages were chosen to capture early childhood, mid-childhood, and late adolescence within the 6–17 year range, corresponding to well-established morphological milestones in cortical maturation and skull development reported in pediatric MRI studies. For each individual, 100 conductivity samples were generated via LHS across five tissue types (scalp, bone, CSF, GM, and WM), yielding 100 perturbed head models. Multi-objective optimization targeting M1 was rerun on each model to construct perturbed Pareto fronts. For Strategy 1 (dose consistency), robustness was defined as whether each perturbed front still contained a solution yielding target-region intensity close to the fixed reference value of 0.15 V/m, thereby demonstrating that dosing consistency could be preserved despite conductivity variability. For Strategy 2 (target engagement), defined as the sparsified 2×2 montage obtained by applying a 0.1 mA threshold, robustness was assessed in two ways: (i) whether mean field norm and target-region intensity remained stable after thresholding, and (ii) whether target-region intensity consistently exceeded that of the conventional montage across perturbations. Together, these analyses evaluated the stability and clinical feasibility of both dose consistency- and target engagement-oriented strategies under large-scale biophysical variability.

In parallel, to quantify how tissue conductivities deform the overall Pareto-front distribution, we computed the Mean Ideal Distance (MID) \cite{garcia2019efficient} for each perturbed front, with the ideal point defined as the solution maximizing target intensity while minimizing whole-brain current spread. We then fit  OLS regression models using the five conductivity parameters as predictors and MID as the response variable \cite{lo2015transform}, enabling identification of the most influential tissues shaping optimization variability. To further assess depth-dependent effects, the same sensitivity analysis was extended to two additional targets—the LDLPFC and hippocampus—representing superficial and deep regions, respectively. These analyses allowed characterization of how the relative influence of tissue conductivities shifts systematically with target depth.

Together, this analysis characterizes both the robustness of strategy selection and the sensitivity of optimization outcomes to tissue conductivity uncertainty, providing a quantitative basis for refining stimulation protocols in heterogeneous populations.

\subsection{Data description}
The T1-weighted MRI data  utilized in this research was acquired from the IPCAS7 internal database, managed by the Institute of Psychology, Chinese Academy of Sciences. This database is part of the International Neuroimaging Data-sharing Initiative (INDI), aimed at facilitating the sharing and standardized processing of neuroimaging data (
\url{https://fcon_1000.projects.nitrc.org/indi/CoRR/html/ipcas_7.html}). After preliminary screening to exclude four participants due to incomplete data or labels, the final sample consisted of 70 children (age range: 6-17 years; 30 males and 40 females). 

All MRI images were captured with a 3T Siemens TrioTim scanner, leveraging an 8-channel head coil for high-fidelity anatomical resolution. We employed a 3D MPRAGE sequence for T1-weighted images, featuring a repetition time (TR) of 2600 ms, an echo time (TE) of 3.02 ms, and a flip angle of 8 degrees. The images are characterized by a slice thickness of 1 mm and an in-plane resolution of 1$\times$1 mm², ensuring a detailed representation of the brain's structure. The field of view (FOV) spans 256 mm, with an acquisition matrix of 256$\times$256, which allows for comprehensive brain coverage without sacrificing detail. This rigorous acquisition protocol ensures a dataset with the precision necessary for advanced neurological studies and applications.

% \backmatter

% \bmhead{Supplementary information}
\section*{Supplementary information}
Supplementary Information accompanies this paper.
It includes: Supplementary Methods, Supplementary Figures, Supplementary Tables, and additional analyses supporting the main text.

\section*{Declarations}

\begin{itemize}
\item Acknowledgements

This work was supported by the National Natural Science Foundation of China (62472206), National Key R\&D Program of China (2025YFC3410000), Shenzhen Science and Technology Innovation Committee (RCYX20231211090405003, KJZD20230923115221044), and the open research fund of the Guangdong Provincial Key Laboratory of Mathematical and Neural Dynamical Systems, the Center for Computational Science and Engineering at Southern University of Science and Technology. WT acknowledges the Pablo J Salame Goldman Sachs Associate Professorship in Computational Neuroscience at Brown University. ZL and YY were supported by American Heart Association Research Award (932980). The authors also thank Chen Wei, Jiahao Tang, Guanyi Zhao, and Zhuo Wang for their valuable assistance of this research.

\item Conflict of interest/Competing interests 

The authors declare no competing interests.

\item Data availability 
The T1-weighted MRI data analyzed in this study were obtained from the IPCAS7 internal database managed by the Institute of Psychology, Chinese Academy of Sciences, which is part of the International Neuroimaging Data-Sharing Initiative (INDI) Consortium for Reliability and Reproducibility (CoRR). Access to the IPCAS7 dataset is subject to the data provider’s requirements and data-use agreements. Public information about the dataset is available at \url{https://fcon_1000.projects.nitrc.org/indi/CoRR/html/ipcas_7.html}. The processed results are publicly available at \url{https://github.com/ncclab-sustech/Pediatric_MOVEA}.
\item Code availability 

	The codes used for head modelling and optimization in this study are publicly available at \url{https://github.com/ncclab-sustech/Pediatric_MOVEA}.
\item Author contribution

	Z.L., M.W., W.T. and Q.L. designed the research; Z.L., M.W. and X.P. conducted the experiments; Z.L., M.W. and X.P. designed the visualization; Z.L., M.W., W.T. and Q.L. wrote the paper; YY reviewed and revised the paper; all the authors analyzed the results; all the authors approved the paper.
\end{itemize}

\bibliography{sn-bibliography}

@article{liang2022online,
  title={Online learning koopman operator for closed-loop electrical neurostimulation in epilepsy},
  author={Liang, Zhichao and Luo, Zixiang and Liu, Keyin and Qiu, Jingwei and Liu, Quanying},
  journal={IEEE Journal of Biomedical and Health Informatics},
  volume={27},
  number={1},
  pages={492--503},
  year={2022},
  publisher={IEEE}
}

@article{moosavi2025controllability,
  title={Controllability of nonlinear epileptic-seizure spreading dynamics in large-scale subject-specific brain networks},
  author={Moosavi, S Amin and Feldman, Jordan S and Truccolo, Wilson},
  journal={Scientific Reports},
  volume={15},
  number={1},
  pages={6467},
  year={2025},
  publisher={Nature Publishing Group UK London}
}

@article{islam2025noninvasive,
  title={Noninvasive brain stimulation as focal epilepsy treatment in the hospital, clinic, and home},
  author={Islam, Karimul and Starnes, Keith and Smith, Kelsey M and Richner, Thomas and Gregg, Nicholas and Rabinstein, Alejandro A and Worrell, Gregory A and Lundstrom, Brian N},
  journal={Epilepsia Open},
  year={2025},
  publisher={Wiley Online Library}
}

@article{lefaucheur2017evidence,
  title={Evidence-based guidelines on the therapeutic use of transcranial direct current stimulation (tDCS)},
  author={Lefaucheur, Jean-Pascal and Antal, Andrea and Ayache, Samar S and Benninger, David H and Brunelin, J{\'e}r{\^o}me and Cogiamanian, Filippo and Cotelli, Maria and De Ridder, Dirk and Ferrucci, Roberta and Langguth, Berthold and others},
  journal={Clinical neurophysiology},
  volume={128},
  number={1},
  pages={56--92},
  year={2017},
  publisher={Elsevier}
}

@article{dasilva2015state,
  title={State-of-art neuroanatomical target analysis of high-definition and conventional tDCS montages used for migraine and pain control},
  author={DaSilva, Alexandre F and Truong, Dennis Q and DosSantos, Marcos F and Toback, Rebecca L and Datta, Abhishek and Bikson, Marom},
  journal={Frontiers in neuroanatomy},
  volume={9},
  pages={89},
  year={2015},
  publisher={Frontiers Media SA}
}

@article{lim2024high,
  title={High-definition tDCS over primary motor cortex modulates brain signal variability and functional connectivity in episodic migraine},
  author={Lim, Manyoel and Kim, Dajung J and Nascimento, Thiago D and DaSilva, Alexandre F},
  journal={Clinical Neurophysiology},
  year={2024},
  publisher={Elsevier}
}

@article{dasilva2022concept,
  title={The concept, development, and application of a home-based high-definition tDCS for bilateral motor cortex modulation in migraine and pain},
  author={DaSilva, Alexandre F and Datta, Abhishek and Swami, Jaiti and Kim, Dajung J and Patil, Parag G and Bikson, Marom},
  journal={Frontiers in Pain Research},
  volume={3},
  pages={798056},
  year={2022},
  publisher={Frontiers Media SA}
}

@article{antonenko2021inter,
  title={Inter-individual and age-dependent variability in simulated electric fields induced by conventional transcranial electrical stimulation},
  author={Antonenko, Daria and Grittner, Ulrike and Saturnino, Guilherme and Nierhaus, Till and Thielscher, Axel and Fl{\"o}el, Agnes},
  journal={Neuroimage},
  volume={224},
  pages={117413},
  year={2021},
  publisher={Elsevier}
}

@article{laakso2015inter,
  title={Inter-subject variability in electric fields of motor cortical tDCS},
  author={Laakso, Ilkka and Tanaka, Satoshi and Koyama, Soichiro and De Santis, Valerio and Hirata, Akimasa},
  journal={Brain stimulation},
  volume={8},
  number={5},
  pages={906--913},
  year={2015},
  publisher={Elsevier}
}

@article{bhattacharjee2022sex,
  title={Sex difference in tDCS current mediated by changes in cortical anatomy: A study across young, middle and older adults},
  author={Bhattacharjee, Sagarika and Kashyap, Rajan and Goodwill, Alicia M and O'Brien, Beth Ann and Rapp, Brenda and Oishi, Kenichi and Desmond, John E and Chen, SH Annabel},
  journal={Brain stimulation},
  volume={15},
  number={1},
  pages={125--140},
  year={2022},
  publisher={Elsevier}
}

@article{mosayebi2021impact,
  title={The impact of individual electrical fields and anatomical factors on the neurophysiological outcomes of tDCS: A TMS-MEP and MRI study},
  author={Mosayebi-Samani, Mohsen and Jamil, Asif and Salvador, Ricardo and Ruffini, Giulio and Haueisen, Jens and Nitsche, Michael A},
  journal={Brain stimulation},
  volume={14},
  number={2},
  pages={316--326},
  year={2021},
  publisher={Elsevier}
}

@article{truong2013computational,
  title={Computational modeling of transcranial direct current stimulation (tDCS) in obesity: impact of head fat and dose guidelines},
  author={Truong, Dennis Q and Magerowski, Greta and Blackburn, George L and Bikson, Marom and Alonso-Alonso, Miguel},
  journal={NeuroImage: Clinical},
  volume={2},
  pages={759--766},
  year={2013},
  publisher={Elsevier}
}

@article{gabriel2009electrical,
  title={Electrical conductivity of tissue at frequencies below 1 MHz},
  author={Gabriel, Camelia and Peyman, Azadeh and Grant, Edwin H},
  journal={Physics in medicine \& biology},
  volume={54},
  number={16},
  pages={4863},
  year={2009},
  publisher={IOP Publishing}
}

@article{opitz2015determinants,
  title={Determinants of the electric field during transcranial direct current stimulation},
  author={Opitz, Alexander and Paulus, Walter and Will, Susanne and Antunes, Andre and Thielscher, Axel},
  journal={Neuroimage},
  volume={109},
  pages={140--150},
  year={2015},
  publisher={Elsevier}
}

@article{mccann2019variation,
  title={Variation in reported human head tissue electrical conductivity values},
  author={McCann, Hannah and Pisano, Giampaolo and Beltrachini, Leandro},
  journal={Brain topography},
  volume={32},
  pages={825--858},
  year={2019},
  publisher={Springer}
}

@article{russell2017sex,
  title={Sex and electrode configuration in transcranial electrical stimulation},
  author={Russell, Michael J and Goodman, Theodore A and Visse, Joseph M and Beckett, Laurel and Saito, Naomi and Lyeth, Bruce G and Recanzone, Gregg H},
  journal={Frontiers in psychiatry},
  volume={8},
  pages={271602},
  year={2017},
  publisher={Frontiers}
}

@article{schmidt2015impact,
  title={Impact of uncertain head tissue conductivity in the optimization of transcranial direct current stimulation for an auditory target},
  author={Schmidt, Christian and Wagner, Sven and Burger, Martin and van Rienen, Ursula and Wolters, Carsten H},
  journal={Journal of neural engineering},
  volume={12},
  number={4},
  pages={046028},
  year={2015},
  publisher={IOP Publishing}
}

@article{shahid2013numerical,
  title={Numerical investigation of white matter anisotropic conductivity in defining current distribution under tDCS},
  author={Shahid, Salman and Wen, Peng and Ahfock, Tony},
  journal={Computer methods and programs in biomedicine},
  volume={109},
  number={1},
  pages={48--64},
  year={2013},
  publisher={Elsevier}
}

@article{jia2022deeptdcs,
  title={DeeptDCS: Deep learning-based estimation of currents induced during transcranial direct current stimulation},
  author={Jia, Xiaofan and Sayed, Sadeed Bin and Hasan, Nahian Ibn and Gomez, Luis J and Huang, Guang-Bin and Yucel, Abdulkadir C},
  journal={IEEE Transactions on Biomedical Engineering},
  volume={70},
  number={4},
  pages={1231--1241},
  year={2022},
  publisher={IEEE}
}

@article{rudroff2020response,
  title={Response variability in transcranial direct current stimulation: why sex matters},
  author={Rudroff, Thorsten and Workman, Craig D and Fietsam, Alexandra C},
  journal={Frontiers in psychiatry},
  volume={11},
  pages={553874},
  year={2020},
  publisher={Frontiers}
}

@article{indahlastari2020modeling,
  title={Modeling transcranial electrical stimulation in the aging brain},
  author={Indahlastari, Aprinda and Albizu, Alejandro and O’Shea, Andrew and Forbes, Megan A and Nissim, Nicole R and Kraft, Jessica N and Evangelista, Nicole D and Hausman, Hanna K and Woods, Adam J and Alzheimer’s Disease Neuroimaging Initiative and others},
  journal={Brain stimulation},
  volume={13},
  number={3},
  pages={664--674},
  year={2020},
  publisher={Elsevier}
}

@article{wang2023multi,
  title={Multi-objective optimization via evolutionary algorithm (MOVEA) for high-definition transcranial electrical stimulation of the human brain},
  author={Wang, Mo and Lou, Kexin and Liu, Zeming and Wei, Pengfei and Liu, Quanying},
  journal={NeuroImage},
  volume={280},
  pages={120331},
  year={2023},
  publisher={Elsevier}
}

@article{saturnino2019principled,
  title={A principled approach to conductivity uncertainty analysis in electric field calculations},
  author={Saturnino, Guilherme B and Thielscher, Axel and Madsen, Kristoffer H and Kn{\"o}sche, Thomas R and Weise, Konstantin},
  journal={Neuroimage},
  volume={188},
  pages={821--834},
  year={2019},
  publisher={Elsevier}
}

@article{saturnino2019simnibs,
  title={SimNIBS 2.1: a comprehensive pipeline for individualized electric field modelling for transcranial brain stimulation},
  author={Saturnino, Guilherme B and Puonti, Oula and Nielsen, Jesper D and Antonenko, Daria and Madsen, Kristoffer H and Thielscher, Axel},
  journal={Brain and human body modeling: computational human modeling at EMBC 2018},
  pages={3--25},
  year={2019},
  publisher={Springer International Publishing}
}

@inproceedings{thielscher2015field,
  title={Field modeling for transcranial magnetic stimulation: A useful tool to understand the physiological effects of TMS?},
  author={Thielscher, Axel and Antunes, Andre and Saturnino, Guilherme B},
  booktitle={2015 37th annual international conference of the IEEE engineering in medicine and biology society (EMBC)},
  pages={222--225},
  year={2015},
  organization={IEEE}
}

@techreport{windhoff2013electric,
  title={Electric field calculations in brain stimulation based on finite elements: an optimized processing pipeline for the generation and usage of accurate individual head models},
  author={Windhoff, Mirko and Opitz, Alexander and Thielscher, Axel},
  year={2013},
  institution={Wiley Online Library}
}

@article{jurcak200710,
  title={10/20, 10/10, and 10/5 systems revisited: their validity as relative head-surface-based positioning systems},
  author={Jurcak, Valer and Tsuzuki, Daisuke and Dan, Ippeita},
  journal={Neuroimage},
  volume={34},
  number={4},
  pages={1600--1611},
  year={2007},
  publisher={Elsevier}
}

@inproceedings{lacadie2008brodmann,
  title={Brodmann Areas defined in MNI space using a new Tracing Tool in BioImage Suite},
  author={Lacadie, CM and Fulbright, RK and Arora, J and Constable, R and Papademetris, X},
  booktitle={Proceedings of the 14th annual meeting of the organization for human brain mapping},
  volume={771},
  year={2008}
}

@article{nielsen2018automatic,
  title={Automatic skull segmentation from MR images for realistic volume conductor models of the head: Assessment of the state-of-the-art},
  author={Nielsen, Jesper D and Madsen, Kristoffer H and Puonti, Oula and Siebner, Hartwig R and Bauer, Christian and Madsen, Camilla G{\o}bel and Saturnino, Guilherme B and Thielscher, Axel},
  journal={Neuroimage},
  volume={174},
  pages={587--598},
  year={2018},
  publisher={Elsevier}
}

@article{saturnino2019electric,
  title={Electric field simulations for transcranial brain stimulation using FEM: an efficient implementation and error analysis},
  author={Saturnino, Guilherme B and Madsen, Kristoffer H and Thielscher, Axel},
  journal={Journal of neural engineering},
  volume={16},
  number={6},
  pages={066032},
  year={2019},
  publisher={IOP Publishing}
}

@article{laakso2012fast,
  title={Fast multigrid-based computation of the induced electric field for transcranial magnetic stimulation},
  author={Laakso, Ilkka and Hirata, Akimasa},
  journal={Physics in Medicine \& Biology},
  volume={57},
  number={23},
  pages={7753},
  year={2012},
  publisher={IOP Publishing}
}

@article{neuling2012finite,
  title={Finite-element model predicts current density distribution for clinical applications of tDCS and tACS},
  author={Neuling, Toralf and Wagner, Sven and Wolters, Carsten H and Zaehle, Tino and Herrmann, Christoph S},
  journal={Frontiers in psychiatry},
  volume={3},
  pages={25619},
  year={2012},
  publisher={Frontiers}
}

@misc{griffiths2018college,
  title={College R (1999) Introduction to electrodynamics},
  author={Griffiths, DJ},
  year={2018},
  publisher={Prentice Hall, New Jersey}
}

@article{miranda2006modeling,
  title={Modeling the current distribution during transcranial direct current stimulation},
  author={Miranda, Pedro Cavaleiro and Lomarev, Mikhail and Hallett, Mark},
  journal={Clinical neurophysiology},
  volume={117},
  number={7},
  pages={1623--1629},
  year={2006},
  publisher={Elsevier}
}

@article{wagner2007transcranial,
  title={Transcranial direct current stimulation: a computer-based human model study},
  author={Wagner, Tim and Fregni, Felipe and Fecteau, Shirley and Grodzinsky, Alan and Zahn, Markus and Pascual-Leone, Alvaro},
  journal={Neuroimage},
  volume={35},
  number={3},
  pages={1113--1124},
  year={2007},
  publisher={Elsevier}
}

@article{wagner2004three,
  title={Three-dimensional head model simulation of transcranial magnetic stimulation},
  author={Wagner, Tim A and Zahn, Markus and Grodzinsky, Alan J and Pascual-Leone, Alvaro},
  journal={IEEE Transactions on Biomedical Engineering},
  volume={51},
  number={9},
  pages={1586--1598},
  year={2004},
  publisher={IEEE}
}

@article{saturnino2015importance,
  title={On the importance of electrode parameters for shaping electric field patterns generated by tDCS},
  author={Saturnino, Guilherme B and Antunes, Andr{\'e} and Thielscher, Axel},
  journal={Neuroimage},
  volume={120},
  pages={25--35},
  year={2015},
  publisher={Elsevier}
}

@article{marino2008methodology,
  title={A methodology for performing global uncertainty and sensitivity analysis in systems biology},
  author={Marino, Simeone and Hogue, Ian B and Ray, Christian J and Kirschner, Denise E},
  journal={Journal of theoretical biology},
  volume={254},
  number={1},
  pages={178--196},
  year={2008},
  publisher={Elsevier}
}

@article{oostendorp2000conductivity,
  title={The conductivity of the human skull: results of in vivo and in vitro measurements},
  author={Oostendorp, Thom F and Delbeke, Jean and Stegeman, Dick F},
  journal={IEEE transactions on biomedical engineering},
  volume={47},
  number={11},
  pages={1487--1492},
  year={2000},
  publisher={IEEE}
}

@article{akhtari2010variable,
  title={Variable anisotropic brain electrical conductivities in epileptogenic foci},
  author={Akhtari, Massoud and Mandelkern, Mark and Bui, Diem and Salamon, Noriko and Vinters, Harry V and Mathern, Gary W},
  journal={Brain topography},
  volume={23},
  pages={292--300},
  year={2010},
  publisher={Springer}
}

@article{li1968specific,
  title={Specific resistivity of the cerebral cortex and white matter},
  author={Li, Choh-luh and Bak, Anthony F and Parker, Levon O},
  journal={Experimental neurology},
  volume={20},
  number={4},
  pages={544--557},
  year={1968},
  publisher={Elsevier}
}

@article{baumann1997electrical,
  title={The electrical conductivity of human cerebrospinal fluid at body temperature},
  author={Baumann, Stephen B and Wozny, David R and Kelly, Shawn K and Meno, Frank M},
  journal={IEEE transactions on biomedical engineering},
  volume={44},
  number={3},
  pages={220--223},
  year={1997},
  publisher={IEEE}
}

@inproceedings{fernandez2016effects,
  title={Effects of head model inaccuracies on regional scalp and skull conductivity estimation using real EIT measurements},
  author={Fern{\'a}ndez-Corazza, Mariano and Turovets, S and Govyadinov, P and Muravchik, CH and Tucker, D},
  booktitle={II Latin American Conference on Bioimpedance: 2nd CLABIO, Montevideo, September 30-October 02, 2015},
  pages={5--8},
  year={2016},
  organization={Springer}
}

@article{fernandez2017skull,
  title={Skull modeling effects in conductivity estimates using parametric electrical impedance tomography},
  author={Fern{\'a}ndez-Corazza, Mariano and Turovets, Sergei and Luu, Phan and Price, Nick and Muravchik, Carlos Horacio and Tucker, Don},
  journal={IEEE Transactions on Biomedical Engineering},
  volume={65},
  number={8},
  pages={1785--1797},
  year={2017},
  publisher={IEEE}
}

@article{bashar2010uncertainty,
  title={Uncertainty and sensitivity analysis for anisotropic inhomogeneous head tissue conductivity in human head modelling},
  author={Bashar, MR and Li, Y and Wen, P},
  journal={Australasian physical \& engineering sciences in medicine},
  volume={33},
  pages={145--152},
  year={2010},
  publisher={Springer}
}

@article{yang2009novel,
  title={A novel strategy of pareto-optimal solution searching in multi-objective particle swarm optimization (MOPSO)},
  author={Yang, Junjie and Zhou, Jianzhong and Liu, Li and Li, Yinghai},
  journal={Computers \& Mathematics with Applications},
  volume={57},
  number={11-12},
  pages={1995--2000},
  year={2009},
  publisher={Elsevier}
}

@article{baron1986moderator,
  title={The moderator--mediator variable distinction in social psychological research: Conceptual, strategic, and statistical considerations.},
  author={Baron, Reuben M and Kenny, David A},
  journal={Journal of personality and social psychology},
  volume={51},
  number={6},
  pages={1173},
  year={1986},
  publisher={American Psychological Association}
}

@article{preacher2008asymptotic,
  title={Asymptotic and resampling strategies for assessing and comparing indirect effects in multiple mediator models},
  author={Preacher, Kristopher J and Hayes, Andrew F},
  journal={Behavior research methods},
  volume={40},
  number={3},
  pages={879--891},
  year={2008},
  publisher={Springer}
}

@article{nikolin2015focalised,
  title={Focalised stimulation using high definition transcranial direct current stimulation (HD-tDCS) to investigate declarative verbal learning and memory functioning},
  author={Nikolin, Stevan and Loo, Colleen K and Bai, Siwei and Dokos, Socrates and Martin, Donel M},
  journal={Neuroimage},
  volume={117},
  pages={11--19},
  year={2015},
  publisher={Elsevier}
}

@article{ciechanski2016transcranial,
  title={Transcranial direct-current stimulation (tDCS): principles and emerging applications in children},
  author={Ciechanski, P and Kirton, A},
  journal={Pediatric Brain Stimulation},
  pages={85--115},
  year={2016},
  publisher={Elsevier}
}

@article{palm2016transcranial,
  title={Transcranial direct current stimulation in children and adolescents: a comprehensive review},
  author={Palm, Ulrich and Segmiller, Felix M and Epple, Ann Natascha and Freisleder, Franz-Joseph and Koutsouleris, Nikolaos and Schulte-K{\"o}rne, Gerd and Padberg, Frank},
  journal={Journal of neural transmission},
  volume={123},
  pages={1219--1234},
  year={2016},
  publisher={Springer}
}

@article{sierawska2023transcranial,
  title={Transcranial direct current stimulation (tDCS) in pediatric populations—--Voices from typically developing children and adolescents and their parents},
  author={Sierawska, Anna and Splittgerber, Maike and Moliadze, Vera and Siniatchkin, Michael and Buyx, Alena},
  journal={Neuroethics},
  volume={16},
  number={1},
  pages={3},
  year={2023},
  publisher={Springer}
}

@article{salehinejad2023optimized,
  title={Optimized HD-tDCS protocol for clinical use in patients with major depressive disorder},
  author={Salehinejad, Mohammad Ali and Abdi, Marzieh and Dadashi, Mohsen and Rostami, Reza and Salvador, Ricardo and Ruffini, Giulio and Nitsche, Michael},
  journal={Brain Stimulation: Basic, Translational, and Clinical Research in Neuromodulation},
  volume={16},
  number={1},
  pages={213},
  year={2023},
  publisher={Elsevier}
}

@article{ngan2022high,
  title={High-definition transcranial direct current stimulation (HD-tDCS) as augmentation therapy in late-life depression (LLD) with suboptimal response to treatment—a study protocol for a double-blinded randomized sham-controlled trial},
  author={Ngan, Sze Ting Joanna and Chan, Lap Kei and Chan, Wai Chi and Lam, Linda Chiu Wa and Li, Wan Kei and Lim, Kelvin and Or, Ego and Pang, Pui Fai and Poon, Ting Keung and Wong, Mei Cheung Mimi and others},
  journal={Trials},
  volume={23},
  number={1},
  pages={914},
  year={2022},
  publisher={Springer}
}

@article{donnell2015high,
  title={High-definition and non-invasive brain modulation of pain and motor dysfunction in chronic TMD},
  author={Donnell, Adam and Nascimento, Thiago D and Lawrence, Mara and Gupta, Vikas and Zieba, Tina and Truong, Dennis Q and Bikson, Marom and Datta, Abhi and Bellile, Emily and DaSilva, Alexandre F},
  journal={Brain stimulation},
  volume={8},
  number={6},
  pages={1085--1092},
  year={2015},
  publisher={Elsevier}
}

@article{villamar2013focal,
  title={Focal modulation of the primary motor cortex in fibromyalgia using 4$\times$ 1-ring high-definition transcranial direct current stimulation (HD-tDCS): immediate and delayed analgesic effects of cathodal and anodal stimulation},
  author={Villamar, Mauricio F and Wivatvongvana, Pakorn and Patumanond, Jayanton and Bikson, Marom and Truong, Dennis Q and Datta, Abhishek and Fregni, Felipe},
  journal={The Journal of Pain},
  volume={14},
  number={4},
  pages={371--383},
  year={2013},
  publisher={Elsevier}
}

@article{buchanan2023safety,
  title={Safety and tolerability of tDCS across different ages, sexes, diagnoses, and amperages: A randomized double-blind controlled study},
  author={Buchanan, Derrick M and Amare, Sarah and Gaumond, Genevieve and D’Angiulli, Amedeo and Robaey, Philippe},
  journal={Journal of Clinical Medicine},
  volume={12},
  number={13},
  pages={4346},
  year={2023},
  publisher={MDPI}
}

@article{andrade2014feasibility,
  title={Feasibility of transcranial direct current stimulation use in children aged 5 to 12 years},
  author={Andrade, Agnes Carvalho and Magnavita, Guilherme Moreira and Allegro, Juleilda Val{\'e}ria Brasil Nunes and Neto, Carlos Eduardo Borges Passos and Lucena, Rita de C{\'a}ssia Saldanha and Fregni, Felipe},
  journal={Journal of child neurology},
  volume={29},
  number={10},
  pages={1360--1365},
  year={2014},
  publisher={SAGE Publications Sage CA: Los Angeles, CA}
}

@article{kessler2013dosage,
  title={Dosage considerations for transcranial direct current stimulation in children: a computational modeling study},
  author={Kessler, Sudha Kilaru and Minhas, Preet and Woods, Adam J and Rosen, Alyssa and Gorman, Casey and Bikson, Marom},
  journal={PloS one},
  volume={8},
  number={9},
  pages={e76112},
  year={2013},
  publisher={Public Library of Science San Francisco, USA}
}

@inproceedings{minhas2012transcranial,
  title={Transcranial direct current stimulation in pediatric brain: a computational modeling study},
  author={Minhas, Preet and Bikson, Marom and Woods, Adam J and Rosen, Alyssa R and Kessler, Sudha K},
  booktitle={2012 annual international conference of the IEEE engineering in medicine and biology society},
  pages={859--862},
  year={2012},
  organization={IEEE}
}

@article{muller2023hd,
  title={HD-tDCS induced changes in resting-state functional connectivity: Insights from EF modeling},
  author={M{\"u}ller, Dario and Habel, Ute and Brodkin, Edward S and Clemens, Benjamin and Weidler, Carmen},
  journal={Brain Stimulation},
  volume={16},
  number={6},
  pages={1722--1732},
  year={2023},
  publisher={Elsevier}
}

@article{yu2017mma,
  title={mma: an R package for mediation analysis with multiple mediators},
  author={Yu, Qingzhao and Li, Bin},
  journal={Journal of Open Research Software},
  volume={5},
  number={1},
  pages={11--11},
  year={2017}
}

@article{yu2023package,
  title={Package ‘mma’},
  author={Yu, Qingzhao and Li, Bin and Yu, Maintainer Qingzhao},
  year={2023}
}

@article{dunn2018chapter,
  title={Chapter 11: Positive continuous data: Gamma and inverse gaussian GLMs},
  author={Dunn, Peter K and Smyth, Gordon K and Dunn, Peter K and Smyth, Gordon K},
  journal={Generalized linear models with examples in R},
  pages={425--456},
  year={2018},
  publisher={Springer}
}

@article{benjamini1995controlling,
  title={Controlling the false discovery rate: a practical and powerful approach to multiple testing},
  author={Benjamini, Yoav and Hochberg, Yosef},
  journal={Journal of the Royal statistical society: series B (Methodological)},
  volume={57},
  number={1},
  pages={289--300},
  year={1995},
  publisher={Wiley Online Library}
}

@article{garcia2019efficient,
  title={An efficient Pareto approach for solving the multi-objective flexible job-shop scheduling problem with regular criteria},
  author={Garc{\'\i}a-Le{\'o}n, Andr{\'e}s Alberto and Dauzere-Peres, St{\'e}phane and Mati, Yazid},
  journal={Computers \& Operations Research},
  volume={108},
  pages={187--200},
  year={2019},
  publisher={Elsevier}
}

@article{lo2015transform,
  title={To transform or not to transform: Using generalized linear mixed models to analyse reaction time data},
  author={Lo, Steson and Andrews, Sally},
  journal={Frontiers in psychology},
  volume={6},
  pages={1171},
  year={2015},
  publisher={Frontiers Media SA}
}

@article{ciechanski2018modeling,
  title={Modeling transcranial direct-current stimulation-induced electric fields in children and adults},
  author={Ciechanski, Patrick and Carlson, Helen L and Yu, Sabrina S and Kirton, Adam},
  journal={Frontiers in human neuroscience},
  volume={12},
  pages={268},
  year={2018},
  publisher={Frontiers Media SA}
}

@article{ma2024mapping,
  title={Mapping the electric field of high-definition transcranial electrical stimulation across the lifespan},
  author={Ma, Weiwei and Wang, Feixue and Yi, Yangyang and Huang, Yu and Li, Xinying and Liu, Yaou and Tu, Yiheng},
  journal={Science Bulletin},
  year={2024},
  publisher={Elsevier}
}

@article{datta2009gyri,
  title={Gyri-precise head model of transcranial direct current stimulation: improved spatial focality using a ring electrode versus conventional rectangular pad},
  author={Datta, Abhishek and Bansal, Varun and Diaz, Julian and Patel, Jinal and Reato, Davide and Bikson, Marom},
  journal={Brain stimulation},
  volume={2},
  number={4},
  pages={201--207},
  year={2009},
  publisher={Elsevier}
}

@article{polania2018studying,
  title={Studying and modifying brain function with non-invasive brain stimulation},
  author={Polan{\'\i}a, Rafael and Nitsche, Michael A and Ruff, Christian C},
  journal={Nature neuroscience},
  volume={21},
  number={2},
  pages={174--187},
  year={2018},
  publisher={Nature Publishing Group US New York}
}

@article{antonenko2018age,
  title={Age-dependent effects of brain stimulation on network centrality},
  author={Antonenko, Daria and Nierhaus, Till and Meinzer, Marcus and Prehn, Kristin and Thielscher, Axel and Ittermann, Bernd and Fl{\"o}el, Agnes},
  journal={NeuroImage},
  volume={176},
  pages={71--82},
  year={2018},
  publisher={Elsevier}
}

@article{miranda2017personalized,
  title={Personalized tDCS stimulation parameters for pediatric subjects},
  author={Miranda, PC},
  journal={Brain Stimulation: Basic, Translational, and Clinical Research in Neuromodulation},
  volume={10},
  number={2},
  pages={525},
  year={2017},
  publisher={Elsevier}
}

@article{datta2011individualized,
  title={Individualized model predicts brain current flow during transcranial direct-current stimulation treatment in responsive stroke patient},
  author={Datta, Abhishek and Baker, Julie M and Bikson, Marom and Fridriksson, Julius},
  journal={Brain stimulation},
  volume={4},
  number={3},
  pages={169--174},
  year={2011},
  publisher={Elsevier}
}

@article{wang2025hd,
  title={HD-tDCS effects on social impairment in autism spectrum disorder with sensory processing abnormalities: a randomized controlled trial},
  author={Wang, Yonglu and Li, Zhijia and Ye, Yupei and Li, Yun and Wei, Ran and Gan, Kaiyan and Qian, Yuxin and Xu, Lingxi and Kong, Yue and Guan, Luyang and others},
  journal={Scientific Reports},
  volume={15},
  number={1},
  pages={9772},
  year={2025},
  publisher={Nature Publishing Group UK London}
}

@article{wang2023randomized,
  title={A randomized, sham-controlled trial of high-definition transcranial direct current stimulation on the right orbital frontal cortex in children and adolescents with attention-deficit hyperactivity disorder},
  author={Wang, Yi-chao and Liu, Jun and Wu, Yan-chun and Wei, Yan and Xie, Hong-jing and Zhang, Tao and Zhang, Zhen},
  journal={Frontiers in Psychiatry},
  volume={14},
  pages={987093},
  year={2023},
  publisher={Frontiers Media SA}
}

@article{krauel2025prefrontal,
  title={Prefrontal Transcranial Direct Current Stimulation in Pediatric Attention-Deficit/Hyperactivity Disorder: A Randomized Clinical Trial},
  author={Krauel, Kerstin and Brauer, Hannah and Breitling-Ziegler, Carolin and Freitag, Christine M and Luckhardt, Christina and M{\"u}hlherr, Andreas and Sch{\"u}tz, Magdalena and Boxhoorn, Sara and Ecker, Christine and Castelo-Branco, Miguel and others},
  journal={JAMA network open},
  volume={8},
  number={2},
  pages={e2460477--e2460477},
  year={2025},
  publisher={American Medical Association}
}

@article{huang2019realistic,
  title={Realistic volumetric-approach to simulate transcranial electric stimulation—ROAST—a fully automated open-source pipeline},
  author={Huang, Yu and Datta, Abhishek and Bikson, Marom and Parra, Lucas C},
  journal={Journal of neural engineering},
  volume={16},
  number={5},
  pages={056006},
  year={2019},
  publisher={IOP Publishing}
}

@article{lipka2021resolving,
  title={Resolving heterogeneity in transcranial electrical stimulation efficacy for attention deficit hyperactivity disorder},
  author={Lipka, Ren{\'e}e and Ahlers, Eike and Reed, Thomas L and Karstens, Malin I and Nguyen, Vu and Bajbouj, Malek and Kadosh, Roi Cohen},
  journal={Experimental Neurology},
  volume={337},
  pages={113586},
  year={2021},
  publisher={Elsevier}
}

@article{rasmussen2021high,
  title={High-definition transcranial direct current stimulation improves delayed memory in Alzheimer’s disease patients: a pilot study using computational modeling to optimize electrode position},
  author={Rasmussen, Ingrid Daae and Boayue, Nya Mehnwolo and Mittner, Matthias and Bystad, Martin and Gr{\o}nli, Ole K and Vangberg, Torgil Riise and Csifcs{\'a}k, G{\'a}bor and Aslaksen, Per M},
  journal={Journal of Alzheimer’s Disease},
  volume={83},
  number={2},
  pages={753--769},
  year={2021},
  publisher={SAGE Publications Sage UK: London, England}
}

@inproceedings{huang2018optimized,
  title={Optimized tDCS for targeting multiple brain regions: an integrated implementation},
  author={Huang, Yu and Thomas, Chris and Datta, Abhishek and Parra, Lucas C},
  booktitle={2018 40th Annual International Conference of the IEEE Engineering in Medicine and Biology Society (EMBC)},
  pages={3545--3548},
  year={2018},
  organization={IEEE}
}

@article{caulfield2020transcranial,
  title={Transcranial electrical stimulation motor threshold can estimate individualized tDCS dosage from reverse-calculation electric-field modeling},
  author={Caulfield, Kevin A and Badran, Bashar W and DeVries, William H and Summers, Philipp M and Kofmehl, Emma and Li, Xingbao and Borckardt, Jeffrey J and Bikson, Marom and George, Mark S},
  journal={Brain Stimulation},
  volume={13},
  number={4},
  pages={961--969},
  year={2020},
  publisher={Elsevier}
}

@article{dmochowski2011optimized,
  title={Optimized multi-electrode stimulation increases focality and intensity at target},
  author={Dmochowski, Jacek P and Datta, Abhishek and Bikson, Marom and Su, Yuzhuo and Parra, Lucas C},
  journal={Journal of neural engineering},
  volume={8},
  number={4},
  pages={046011},
  year={2011},
  publisher={IOP Publishing}
}

@article{ruffini2014optimization,
  title={Optimization of multifocal transcranial current stimulation for weighted cortical pattern targeting from realistic modeling of electric fields},
  author={Ruffini, Giulio and Fox, Michael D and Ripolles, Oscar and Miranda, Pedro Cavaleiro and Pascual-Leone, Alvaro},
  journal={Neuroimage},
  volume={89},
  pages={216--225},
  year={2014},
  publisher={Elsevier}
}

@article{kim2024exploring,
  title={Exploring HD-tDCS Effect on $\mu$-opioid Receptor and Pain Sensitivity in Temporomandibular Disorder: A Pilot Randomized Clinical Trial Study},
  author={Kim, Dajung J and Nascimento, Thiago D and Lim, Manyoel and Danciu, Theodora and Zubieta, Jon-Kar and Scott, Peter JH and Koeppe, Robert and Kaciroti, Niko and DaSilva, Alexandre F},
  journal={The journal of pain},
  volume={25},
  number={4},
  pages={1070--1081},
  year={2024},
  publisher={Elsevier}
}

@article{bikson2012computational,
  title={Computational models of transcranial direct current stimulation},
  author={Bikson, Marom and Rahman, Asif and Datta, Abhishek},
  journal={Clinical EEG and neuroscience},
  volume={43},
  number={3},
  pages={176--183},
  year={2012},
  publisher={SAGE Publications Sage CA: Los Angeles, CA}
}

@article{caiani2025anatomical,
  title={Anatomical Characteristics Predict Response to Transcranial Direct Current Stimulation (tDCS): Development of a Computational Pipeline for Optimizing tDCS Protocols},
  author={Caiani, Giulia and Chiaramello, Emma and Parazzini, Marta and Arrigoni, Eleonora and Lauro, Leonor J Romero and Pisoni, Alberto and Fiocchi, Serena},
  journal={Bioengineering},
  volume={12},
  number={6},
  pages={656},
  year={2025},
  publisher={MDPI}
}

@article{bardhi2025optimization,
  title={Optimization Simulations of Transcranial Direct Current Stimulation Montages in Children With Perinatal Stroke},
  author={Bardhi, Martin and Zewdie, Ephrem Takele and Kirton, Adam and Carlson, Helen L},
  journal={Neuromodulation: Technology at the Neural Interface},
  year={2025},
  publisher={Elsevier}
}

@article{koolschijn2013sex,
  title={Sex differences and structural brain maturation from childhood to early adulthood},
  author={Koolschijn, P C{\'e}dric MP and Crone, Eveline A},
  journal={Developmental cognitive neuroscience},
  volume={5},
  pages={106--118},
  year={2013},
  publisher={Elsevier}
}

@inproceedings{rampersad2011handling,
  title={On handling the layered structure of the skull in transcranial direct current stimulation models},
  author={Rampersad, Sumientra and Stegeman, Dick and Oostendorp, Thom},
  booktitle={2011 Annual International Conference of the IEEE Engineering in Medicine and biology Society},
  pages={1989--1992},
  year={2011},
  organization={IEEE}
}

@article{feng2018transcranial,
  title={Transcranial direct current stimulation for poststroke motor recovery: challenges and opportunities},
  author={Feng, Wuwei and Kautz, Steven A and Schlaug, Gottfried and Meinzer, Caitlyn and George, Mark S and Chhatbar, Pratik Y},
  journal={PM\&R},
  volume={10},
  number={9},
  pages={S157--S164},
  year={2018},
  publisher={Elsevier}
}

@article{evans2020dose,
  title={Dose-controlled tDCS reduces electric field intensity variability at a cortical target site},
  author={Evans, Carys and Bachmann, Clarissa and Lee, Jenny SA and Gregoriou, Evridiki and Ward, Nick and Bestmann, Sven},
  journal={Brain stimulation},
  volume={13},
  number={1},
  pages={125--136},
  year={2020},
  publisher={Elsevier}
}

@article{guler2016optimization,
  title={Optimization of focality and direction in dense electrode array transcranial direct current stimulation (tDCS)},
  author={Guler, Seyhmus and Dannhauer, Moritz and Erem, Burak and Macleod, Rob and Tucker, Don and Turovets, Sergei and Luu, Phan and Erdogmus, Deniz and Brooks, Dana H},
  journal={Journal of neural engineering},
  volume={13},
  number={3},
  pages={036020},
  year={2016},
  publisher={IOP Publishing}
}

@inproceedings{guler2016optimizing,
  title={Optimizing stimulus patterns for dense array tDCS with fewer sources than electrodes using a branch and bound algorithm},
  author={Guler, Seyhmus and Dannhauer, Moritz and Erem, Burak and Macleod, Rob and Tucker, Don and Turovets, Sergei and Luu, Phan and Meleis, Waleed and Brooks, Dana H},
  booktitle={2016 IEEE 13th International Symposium on Biomedical Imaging (ISBI)},
  pages={229--232},
  year={2016},
  organization={IEEE}
}

@article{helton2003latin,
  title={Latin hypercube sampling and the propagation of uncertainty in analyses of complex systems},
  author={Helton, Jon C and Davis, Freddie Joe},
  journal={Reliability Engineering \& System Safety},
  volume={81},
  number={1},
  pages={23--69},
  year={2003},
  publisher={Elsevier}
}

@article{fabbrizzi2025reconstructing,
  title={Reconstructing whole-brain structure and dynamics using imaging data and personalized modeling},
  author={Fabbrizzi, M and Amato, LG and Martinelli, L and Carpaneto, J and Bartolini, E and Calderoni, S and Retico, A and Vergani, AA and Mazzoni, A},
  journal={medRxiv},
  pages={2025--01},
  year={2025},
  publisher={Cold Spring Harbor Laboratory Press}
}

@article{shafiei2025reproducible,
  title={Reproducible Brain Charts: An open data resource for mapping brain development and its associations with mental health},
  author={Shafiei, Golia and Esper, Nathalia B and Hoffmann, Mauricio S and Ai, Lei and Chen, Andrew A and Cluce, Jon and Covitz, Sydney and Giavasis, Steven and Lane, Connor and Mehta, Kahini and others},
  journal={Neuron},
  year={2025},
  publisher={Elsevier}
}

@article{luo2025frequency,
  title={Frequency-specific and state-dependent neural responses to brain stimulation},
  author={Luo, Huichun and Ye, Xiaolai and Cai, Hui-Ting and Wang, Mo and Wang, Yue and Liu, Qiangqiang and Xu, Ying and Mao, Ziyu and Cai, Yanqing and Hong, Jing and others},
  journal={Molecular Psychiatry},
  pages={1--11},
  year={2025},
  publisher={Nature Publishing Group UK London}
}

@article{wang2025computational,
  title={Computational Modeling for Personalized Transcranial Electrical Stimulation: Theory, Tools, and Applications},
  author={Wang, Mo and Zheng, Kexin and Liu, Yiling and Luo, Huichun and Yuan, Tifei and Wen, Hongkai and Wei, Pengfei and Liu, Quanying},
  journal={arXiv preprint arXiv:2509.01192},
  year={2025}
}

@article{williamson2025neuroengineering,
  title={Neuroengineering approaches assessing structural and functional changes of motor descending pathways in stroke},
  author={Williamson, Jordan N and Peng, Rita Huan-Ting and Sung, Joohwan and Darvish, Mahmood Rajabtabar and Chen, Xiaoxi and Ali, Mehreen and Li, Sheng and Yang, Yuan},
  journal={Progress in Biomedical Engineering},
  volume={7},
  number={4},
  pages={042006},
  year={2025},
  publisher={IOP Publishing}
}

@article{williamson2023high,
  title={High-definition transcranial direct current stimulation for upper extremity rehabilitation in moderate-to-severe ischemic stroke: a pilot study},
  author={Williamson, Jordan N and James, Shirley A and He, Dorothy and Li, Sheng and Sidorov, Evgeny V and Yang, Yuan},
  journal={Frontiers in Human Neuroscience},
  volume={17},
  pages={1286238},
  year={2023},
  publisher={Frontiers Media SA}
}

@article{peng2024determining,
  title={Determining the effects of targeted high-definition transcranial direct current stimulation on reducing post-stroke upper limb motor impairments—a randomized cross-over study},
  author={Peng, Rita Huan-Ting and He, Dorothy and James, Shirley A and Williamson, Jordan N and Skadden, Carly and Jain, Sanjiv and Hassaneen, Wael and Miranpuri, Amrendra and Kaur, Amandeep and Sarol, Jesus N and others},
  journal={Trials},
  volume={25},
  number={1},
  pages={34},
  year={2024},
  publisher={Springer}
}
\newpage
\appendix

\renewcommand{\thefigure}{S\arabic{figure}}
\setcounter{figure}{0}
\renewcommand{\thetable}{S\arabic{table}}
\setcounter{table}{0}

\section{Appendix}

\subsection{Limitations of Conventional Montages in Pediatric HD-tDCS}

\begin{figure}[t]
   \includegraphics[width=1\textwidth]{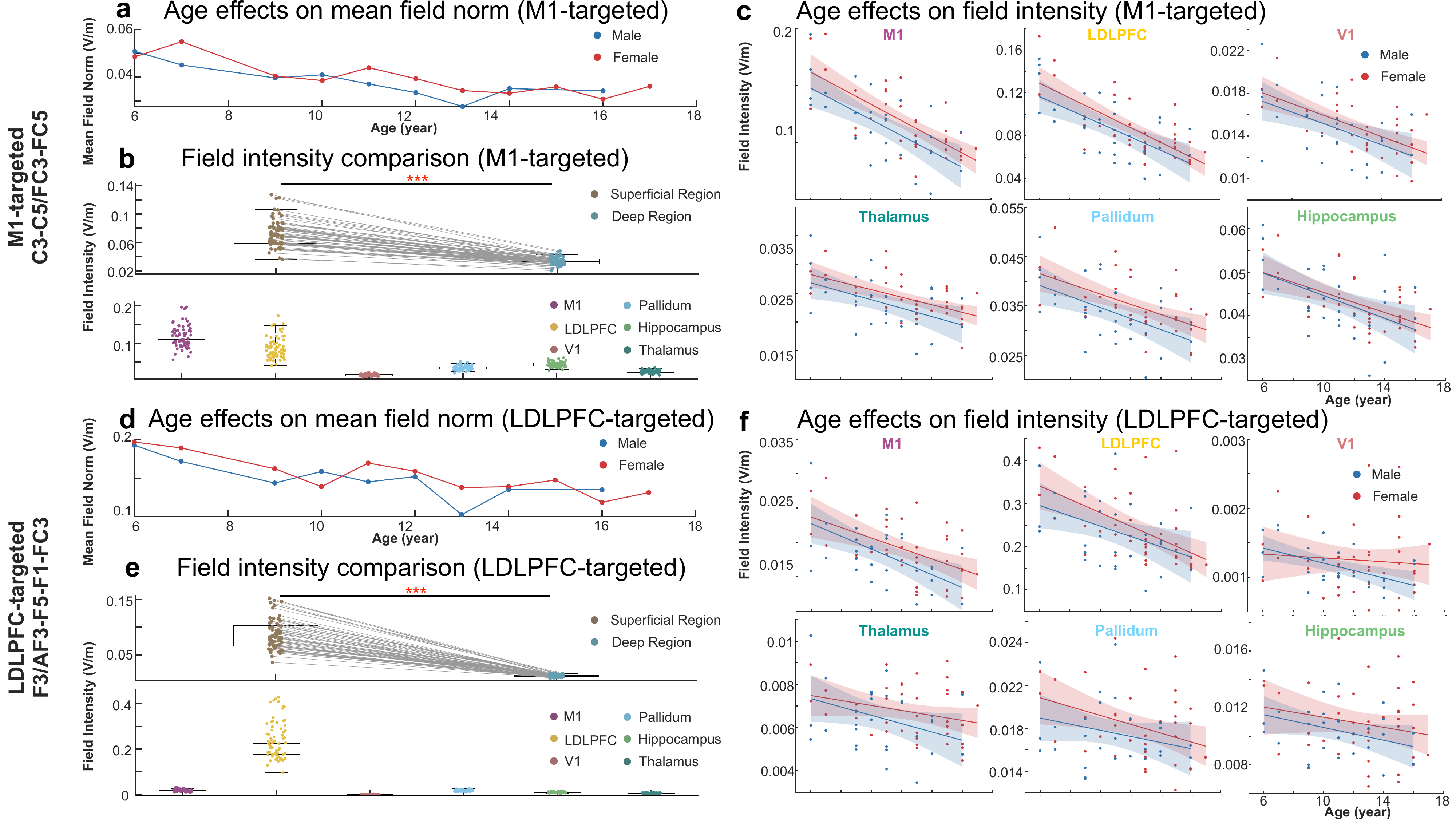}
   \caption{\textbf{Age and Sex Effects on Electric Field Distribution Across Pediatric HD-tDCS Montages}
(\textbf{A}) \textit{Age effects on mean field norm (Migraine treatment)}: The mean electric field intensity norm shows a decreasing trend with age across both sexes, suggesting that younger participants experience stronger electric fields, which decline as age increases. This trend aligns with the findings presented in Fig. 2. 
(\textbf{B}) \textit{Field intensity comparison (Migraine treatment)}: Comparison of field intensity in superficial (M1, V1, LDLPFC) and deep regions (hippocampus, pallidum, thalamus). Significant differences (*** p < 0.001) are observed between superficial and deep regions. 
(\textbf{C}) \textit{Age effects on field intensity (Migraine treatment)}: Field intensity decreases with age across six regions of interest (M1, V1, LDLPFC, hippocampus, pallidum, thalamus), all showing similar trends. M1 and hippocampus, as observed in Fig. 2. \\
(\textbf{D}) \textit{Age effects on mean field norm (Cognitive enhancement)}: Similar to the migraine treatment montage, there is a decreasing trend in mean field norm with age, consistent with Fig. 2.
(\textbf{E}) \textit{Field intensity comparison (Cognitive enhancement)}: Comparison of field intensity in superficial and deep regions for the cognitive enhancement montage, with higher intensities observed in superficial regions. 
(\textbf{F}) \textit{Age effects on field intensity (Cognitive enhancement)}: Field intensity decreases with age in six regions of interest (M1, V1, LDLPFC, hippocampus, pallidum, thalamus), showing similar trends. LDLPFC and hippocampus, as shown in Fig. 2., exhibit consistent patterns of change.
   \label{fig:S1}}
 \end{figure}

Fig.~\ref{fig:S1} extends the main text results by showing age- and sex-dependent electric field distributions across six ROIs for both conventional montages. Consistent with the main text, whole-brain mean field norm decreased with age under both montages (Fig.~\ref{fig:S1} A, D), indicating stronger stimulation in younger children and progressive attenuation during adolescence. This decline was observed across all ROIs, with both superficial (M1, V1, LDLPFC) and deep (hippocampus, pallidum, thalamus) regions showing reductions in field intensity with increasing age. Visual inspection of Panels C and F further suggests that females tended to show slightly higher field intensities than males across most ROIs. In addition, superficial ROIs exhibited systematically higher intensities than deep regions (Fig.~\ref{fig:S1} B, E). These findings underscore that conventional fixed montages fail to provide age-independent dosing.

In line with these visual patterns, nested GLMs analyses provided quantitative evidence for robust age or sex-related effects across nearly all ROIs under both montages (Tables.~\ref{tab:GLM_M1_main},~\ref{tab:GLM_LDLPFC_main},~\ref{tab:GLM_M1_Supplementary}, and~\ref{tab:GLM_LDLPFC_Supplementary}). Tables.~\ref{tab:GLM_M1_Supplementary}--\ref{tab:GLM_LDLPFC_Supplementary} extend the main text analyses by providing statistical results for the remaining ROIs not shown in Tables.~\ref{tab:GLM_M1_main}--\ref{tab:GLM_LDLPFC_main}. When targeting M1, Table.~\ref{tab:GLM_M1_Supplementary} showed that electric field intensity at other four ROIs—including both superficial (V1, LDLPFC) and deep (pallidum, thalamus) as well as mean field norm across whole brain—showed significant age effects (all $p<0.05$, FDR-corrected), reinforcing that developmental stage exerts a pervasive influence on stimulation strength under this montage.
By contrast, Table.~\ref{tab:GLM_LDLPFC_Supplementary} showed that LDLPFC-targeted stimulation exhibited more complex demographic dependencies. In addition to robust age effects observed across all ROIs, sex-related differences emerged at several regions, most notably at M1, V1, and thalamus, and a notable age–sex interaction at V1.

Together, these extended analyses confirm that both conventional montages not only produce stronger age dependence but also introduce broader sex-related variability, thereby highlighting the limitations of conventional fixed strategies in ensuring demographic-independent dosing.

\clearpage   
\begin{table}[H]
\centering
\caption{Age and sex effects on electric field intensity under conventional M1-targeted montage (FC3–FC5/C3–C5).}

\begin{adjustbox}{max width=\textwidth}
\begin{tabular}{ccccccccccc}
\toprule
 & $\beta_0$ & $\beta_1$ & $\beta_2$ & $\beta_3$ & $\beta_4$ &  &  & $\Delta$ Deviance & \textit{p}-value & \textit{p}ID \\
\hline
\rule{0pt}{3pt} \\
\textbf{Primary Motor Cortex} &  &  &  &  &  &  &  &  &  &  \\

$M_0$     & 73.7872 &  &  &  &  &  &  &  &  &  \\
$M_1$    & -5.9831 & 7.3095 &  &  &  &  &  & 15.3596 & 0.0001 & * \\
$M_2$    & -17.6039 & 7.845859 & 13.4564 &  &  &  &  & 1.2754 & 0.2587 &  \\
$M_3$    & -13.5019 & 7.4623 & 2.4571 & 1.1391 &  &  &  & 0.0864 & 0.76884 &  \\
$M_4$   & -13.5019 & 7.4623 & 2.4571 & 1.1391 &0  &  &  & 0 & 1 &  \\

\hline
\rule{0pt}{3pt} \\ 
\textbf{Hippocampus} &  &  &  &  &  &  &  &  &  &  \\
$M_0$    & 539.9094  &  &  &  &  &  &  &  &  &  \\
$M_1$    & 221.7001 & 28.1812 &  &  &  &  &  & 10.8507 & 0.0010 & * \\
$M_2$  & 201.4482 & 29.1745 & 20.4851 &  &  &  &  & 0.1346 & 0.7137 &  \\
$M_3$   & 215.7112 & 27.9636 & -11.1953 & 2.9842 &  &  &  & 0.0275 & 0.8683 &  \\
$M_4$    & 215.7112 & 27.9636 & -11.1953 & 2.9842 & 0 &  &  & 0 & 1 &  \\

\bottomrule

\end{tabular}
\end{adjustbox}

\begin{flushleft}

Nested generalized linear models (GLMs) were used to assess the effects of age, sex, and their interaction on stimulation-driven electric field intensity in two regions of interest (ROIs) highlighted in the main text when targeting the primary motor cortex (M1): the M1 and hippocampus. Model coefficients ($\beta_0$–$\beta_4$) correspond to the intercept, age, sex, and age–sex interaction terms; $\Delta$ Deviance and associated \textit{p}-values indicate model improvements relative to the preceding model ($\chi^2$ test). The \textit{p}ID column denotes significance after false discovery rate (FDR) correction; asterisks indicate $p < 0.05$.

\end{flushleft}
\label{tab:GLM_M1_main}
\end{table}

\begin{table}[H]
\centering
\caption{Age and sex effects on stimulation-driven electric field intensity under conventional montage targeting LDLPFC (F3/AF3–F5–F1–FC3)}
\begin{adjustbox}{max width=\textwidth}
\begin{tabular}{ccccccccccc}

\toprule
 & $\beta_0$ & $\beta_1$ & $\beta_2$ & $\beta_3$ & $\beta_4$ &  &  & $\Delta$ Deviance & \textit{p}-value & \textit{p}ID \\

\hline
\rule{0pt}{3pt} \\ 
\textbf{LDLPFC} &  &  &  &  &  &  &  &  &  &  \\
$M_0$     & 17.3370  &  &  &  &  &  &  &  &  &  \\
$M_1$    & -1.0057 & 1.6778 &  &  &  &  &  & 7.0826 & 0.0078 & * \\
$M_2$    & -3.6624 & 1.8010 & 3.0501 &  &  &  &  & 0.5714  & 0.4497 &  \\
$M_3$    & -4.2272 & 1.8540 & 4.5907 & -0.1583 &  &  &  & 0.0147& 0.9036 &  \\
$M_4$    & -4.2272 & 1.8540 & 4.5907 & -0.1583 & 0  &  &  & 0& 1 &  \\

\hline
\rule{0pt}{3pt} \\ 
\textbf{Hippocampus} &  &  &  &  &  &  &  &  &  &  \\
$M_0$    & 8695.7986  &  &  &  &  &  &  &  &  &  \\
$M_1$    & 5859.6617 & 247.2881 &  &  &  &  &  & 12.5870 & 0.0004 & * \\
$M_2$     & 4636.9518 & 308.6188& 1251.8195&  &  &  &  & 7.5098 & 0.0061 &  *\\
$M_3$    & 5199.4703 & 261.9698 & -112.5041 & 126.3241 &  &  &  & 0.7116  &0.3989 &  \\
$M_4$     & 5199.4703 & 261.9698 & -112.5041 & 126.3241 & 0 &  &  & 0  &1 &  \\

\bottomrule
\end{tabular}
\end{adjustbox}
\begin{flushleft} 

Nested GLMs were used to assess the effects of age, sex, and their interaction on stimulation-driven electric field intensity in two regions of interest (ROIs) highlighted in the main text when targeting the left dorsolateral prefrontal cortex (LDLPFC): the LDLPFC and hippocampus. Model coefficients ($\beta_0$–$\beta_4$) correspond to the intercept, age, sex, and age–sex interaction terms; $\Delta$ Deviance and associated \textit{p}-values indicate model improvements relative to the preceding model ($\chi^2$ test). The \textit{p}ID column denotes significance after false discovery rate (FDR) correction; asterisks indicate $p < 0.05$. 
\end{flushleft}
\label{tab:GLM_LDLPFC_main}
\end{table}

\begin{table}[H]
\centering

\caption{Age and sex effects on electric field intensity under conventional M1-targeted montage (FC3–FC5/C3–C5).}

\begin{adjustbox}{max width=\textwidth}
\begin{tabular}{ccccccccccc}
\toprule
 & $\beta_0$ & $\beta_1$ & $\beta_2$ & $\beta_3$ & $\beta_4$ &  &  & $\Delta$ Deviance & \textit{p}-value & \textit{p}ID \\
\hline
\rule{0pt}{3pt} \\ 
\textbf{LDLPFC} &  &  &  &  &  &  &  &  &  &  \\
$M_0$     & 135.4024  &  &  &  &  &  &  &  &  &  \\
$M_1$   & -47.5876 & 17.1508 &  &  &  &  &  & 35.3980 & 0 & * \\
$M_2$     & -64.2313 & 17.8882 & 19.8370 &  &  &  &  & 1.2189 & 0.2696 &  \\
$M_3$    & -63.9616 & 17.8613 & 19.1087 & 0.0792 &  &  &  & 0.0002 & 0.9892 &  \\
$M_4$  & -63.9616 & 17.8613 & 19.1087 & 0.0792 & 0 &  &  & 0 & 1 &  \\

\hline
\rule{0pt}{3pt} \\  
\textbf{V1} &  &  &  &  &  &  &  &  &  &  \\
$M_0$    & 4565.0240  &  &  &  &  &  &  &  &  &  \\
$M_1$     & 1441.5150 & 278.2914 &  &  &  &  &  & 43.4969  & 0 & * \\
$M_2$    & 1033.0679 & 298.1980 & 426.0099 &  &  &  &  & 2.4011  & 0.1212 &  \\
$M_3$    & 1096.0157 & 292.7513 & 277.5181 & 14.2642 &  &  &  & 0.0259  & 0.8722 &  \\
$M_4$    & 1096.0157 & 292.7513 & 277.5181 & 14.2642 & 0 &  &  & 0  & 1 &  \\

\hline
\rule{0pt}{3pt} \\ 
\textbf{Thalamus} &  &  &  &  &  &  &  &  &  &  \\
$M_0$     & 1770.8828 &  &  &  &  &  &  &  &  &  \\
$M_1$     & 807.3178 & 85.0972 &  &  &  &  &  & 16.5751  & 0 & * \\
$M_2$     & 559.6070 & 97.3195 & 259.8068 &  &  &  &  & 3.6170 & 0.0572 &  \\
$M_3$    & 669.1027 & 87.9975 & -8.8046 & 25.5267&  &  &  & 0.3292 & 0.5661 &  \\
$M_4$     & 669.1027 & 87.9975 & -8.8046 & 25.5267& 0 &  &  & 0 & 1 &  \\

\hline
\rule{0pt}{3pt} \\  
\textbf{Pallidum} &  &  &  &  &  &  &  &  &  &  \\
$M_0$    & 842.6963 &  &  &  &  &  &  &  &  &  \\
$M_1$    & 352.8506 & 43.3598 &  &  &  &  &  & 13.1615 & 0.0003 & * \\
$M_2$    & 230.1719 & 49.3938 & 129.6019 &  &  &  &  & 2.7562  & 0.0969 &  \\
$M_3$     & 279.2402 & 45.1921 & 8.0332 & 11.6171 &  &  &  & 0.2090 & 0.6475 &  \\
$M_4$     & 279.2402 & 45.1921 & 8.0332 & 11.6171 & 0 &  &  & 0 & 1 &  \\

\hline
\rule{0pt}{3pt} \\  
\textbf{Mean} &  &  &  &  &  &  &  &  &  &  \\
$M_0$     & 672.1490 &  &  &  &  &  &  &  &  &  \\
$M_1$     & 76.2744 & 53.8206 &  &  &  &  &  & 29.5094 & 0& * \\
$M_2$     & 2.5834 & 57.3390 & 79.6957 &  &  &  &  & 1.5446 & 0.2139 &  \\
$M_3$     & 26.3171 & 55.2181 & 21.7991 & 5.7500 &  &  &  & 0.0775  & 0.7808 &  \\
$M_4$     & 26.3171 & 55.2181 & 21.7991 & 5.7500 & 0 &  &  & 0 & 1 &  \\

\bottomrule
\end{tabular}
\end{adjustbox}

\begin{flushleft}
Nested GLMs analysis of additional ROIs beyond those reported in the main text when targeting M1, including LDLPFC, V1, thalamus, pallidum, and the mean field norm across whole brain. Model coefficients ($\beta_0$–$\beta_4$) correspond to the intercept, age, sex, and age–sex interaction terms; $\Delta$ Deviance and \textit{p}-values indicate model improvements relative to the preceding model ($\chi^2$ test). The \textit{p}ID column denotes significance after false discovery rate (FDR) correction; asterisks indicate $p < 0.05$. Results for M1 and hippocampus are shown in Table S1.
\end{flushleft}
\label{tab:GLM_M1_Supplementary}
\end{table}

\begin{table}[H]
\centering
\caption{Age and sex effects on stimulation-driven electric field intensity under conventional montage targeting LDLPFC (F3/AF3–F5–F1–FC3)}
\begin{adjustbox}{max width=\textwidth}
\begin{tabular}{ccccccccccc}
\toprule
 & $\beta_0$ & $\beta_1$ & $\beta_2$ & $\beta_3$ & $\beta_4$ &  &  & $\Delta$ Deviance & \textit{p}-value & \textit{p}ID \\
\hline
\rule{0pt}{3pt} \\  
\textbf{Primary Motor Cortex} &  &  &  &  &  &  &  &  &  &  \\
$M_0$     & 2882.5971 &  &  &  &  &  &  &  &  &  \\
$M_1$    & 499.9240 & 214.3090&  &  &  &  &  & 52.2877 & 0 & * \\
$M_2$     & 45.5853 & 236.0801 & 500.8512 &  &  &  &  & 6.7950 & 0.0091 &  *\\
$M_3$     & 387.6554 & 205.7862 & -366.5113& 86.1185 &  &  &  & 1.9000  & 0.1681 &  \\
$M_4$     & 387.6554 & 205.7862 & -366.5113& 86.1185 & 0 &  &  & 0 & 1 &  \\

\hline
\rule{0pt}{3pt} \\
\textbf{V1} &  &  &  &  &  &  &  &  &  &  \\
$M_0$     & 674164.1832 &  &  &  &  &  &  &  &  &  \\
$M_1$     & 378055.2398 & 25987.0154 &  &  &  &  &  & 205.8669  & 0 & * \\
$M_2$     & 247261.7205 & 32504.9350 & 138624.9005 &  &  &  &  & 136.2077   & 0 & *  \\
$M_3$    & 471565.2267 & 13847.1057 & -412402.6191 & 52385.4840 &  &  &  & 178.0784  & 0 & * \\
$M_4$     & 471565.2267 & 13847.1057 & -412402.6191 & 52385.4840 & 0 &  &  & 0  & 1 &  \\

\hline
\rule{0pt}{3pt} \\  
\textbf{Thalamus} &  &  &  &  &  &  &  &  &  &  \\
$M_0$     & 22823.5208 &  &  &  &  &  &  &  &  &  \\
$M_1$    & 13708.3860& 798.1016 &  &  &  &  &  & 31.0497  & 0 & * \\
$M_2$    & 10095.1938 & 978.5395 & 3753.8644 &  &  &  &  & 15.9950 & 0 & * \\
$M_3$     & 13198.2160 & 719.6684 & -3840.6581 & 711.8518&  &  &  & 5.3386 & 0.0209 &  \\
$M_4$     & 13198.2160 & 719.6684 & -3840.6581 & 711.8518& 0 &  &  & 0& 1&  \\

\hline
\rule{0pt}{3pt} \\ 
\textbf{Pallidum} &  &  &  &  &  &  &  &  &  &  \\
$M_0$     & 3102.1001&  &  &  &  &  &  &  &  &  \\
$M_1$     & 1967.9617& 99.1098 &  &  &  &  &  & 9.5247 & 0.0020 & * \\
$M_2$    & 1569.6786 & 119.0222 & 407.5965 &  &  &  &  &3.7564 & 0.0526 &  \\
$M_3$    & 1460.7759 & 128.1346 & 673.5879 & -24.6526 &  &  &  & 0.1293 & 0.7192 &  \\
$M_4$     & 1460.7759 & 128.1346 & 673.5879 & -24.6526 & 0 &  &  & 0 & 1 &  \\

\hline
\rule{0pt}{3pt} \\  
\textbf{Mean} &  &  &  &  &  &  &  &  &  &  \\
$M_0$     & 4409.6377 &  &  &  &  &  &  &  &  &  \\
$M_1$     & 1026.7191 & 303.0523 &  &  &  &  &  & 54.8631 & 0& * \\
$M_2$     & 424.5513 & 332.1406& 649.7038 &  &  &  &  & 5.9708 & 0.0145 &  \\
$M_3$     & 607.1158 & 316.0799 &193.7473 & 44.6429 &  &  &  & 0.2689 &0.6041 &  \\
$M_4$     & 607.1158 & 316.0799 &193.7473 & 44.6429 & 0 &  &  & 0 &1 &  \\

\bottomrule
\end{tabular}
\end{adjustbox}

\begin{flushleft}
Nested GLMs analysis of additional ROIs beyond those reported in the main text when targeting LDLPFC, including M1, V1, thalamus, pallidum, and the mean field norm across whole brain. Model coefficients ($\beta_0$–$\beta_4$) correspond to the intercept, age, sex, and age–sex interaction terms; $\Delta$ Deviance and \textit{p}-values indicate model improvements relative to the preceding model ($\chi^2$ test). The \textit{p}ID column denotes significance after false discovery rate (FDR) correction; asterisks indicate $p < 0.05$. Results for LDLPFC and hippocampus are shown in Table S2.

\end{flushleft} 
\label{tab:GLM_LDLPFC_Supplementary}
\end{table}

\newpage

\subsection{Inverse Dual-Objective Optimization Framework for Pediatric HD-tDCS}

\begin{figure}[htbp]
  \includegraphics[width=1\textwidth]{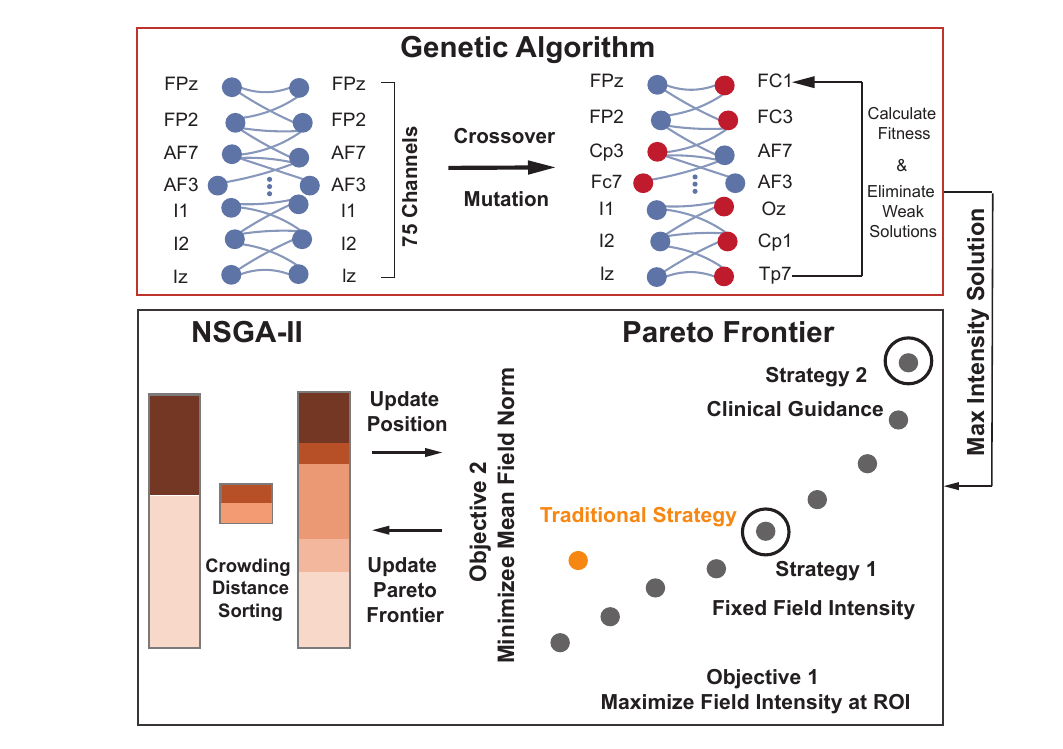}
  \caption{\textbf{Schematic of the Inverse Dual-Objective Optimization Framework for Pediatric HD-tDCS}
  This figure illustrates the process of deriving Pareto frontiers using the inverse dual-objective optimization framework. The framework combines Genetic Algorithm (GA) and NSGA-II (Non-dominated Sorting Genetic Algorithm II) to optimize electrode configurations for pediatric HD-tDCS. The GA is used to generate and evolve potential solutions, represented by 75 electrode positions (blue circles), with crossover and mutation operations guiding the search for optimal solutions. The fitness of each solution is evaluated based on two objectives: (1) maximizing field intensity at the region of interest (ROI) and (2) minimizing the mean field norm. NSGA-II then sorts and updates the solutions to form the Pareto frontiers (grey dots), representing the trade-offs between the two objectives. The final step involves selecting two representative strategies: Strategy 1, which enforces fixed field intensity at M1 for consistent dosing, and Strategy 2, which maximizes M1 field intensity for improved stimulation target engagement. These strategies offer distinct optimization pathways for personalized pediatric HD-tDCS treatment.
  \label{fig:S2}}
\end{figure}

Fig.~\ref{fig:S2} illustrates the schematic of the inverse dual-objective optimization framework, combining Genetic Algorithm (GA) search with NSGA-II sorting to generate subject-specific Pareto fronts. Candidate electrode montages were evaluated according to two competing objectives: maximizing field intensity at target region and minimizing whole-brain field spread. From these Pareto fronts, two representative solutions were derived: Strategy~1 (fixed field intensity for dose consistency) and Strategy~2 (maximal field intensity for target engagement).

\begin{figure}[htbp]
  \includegraphics[width=0.98\textwidth]{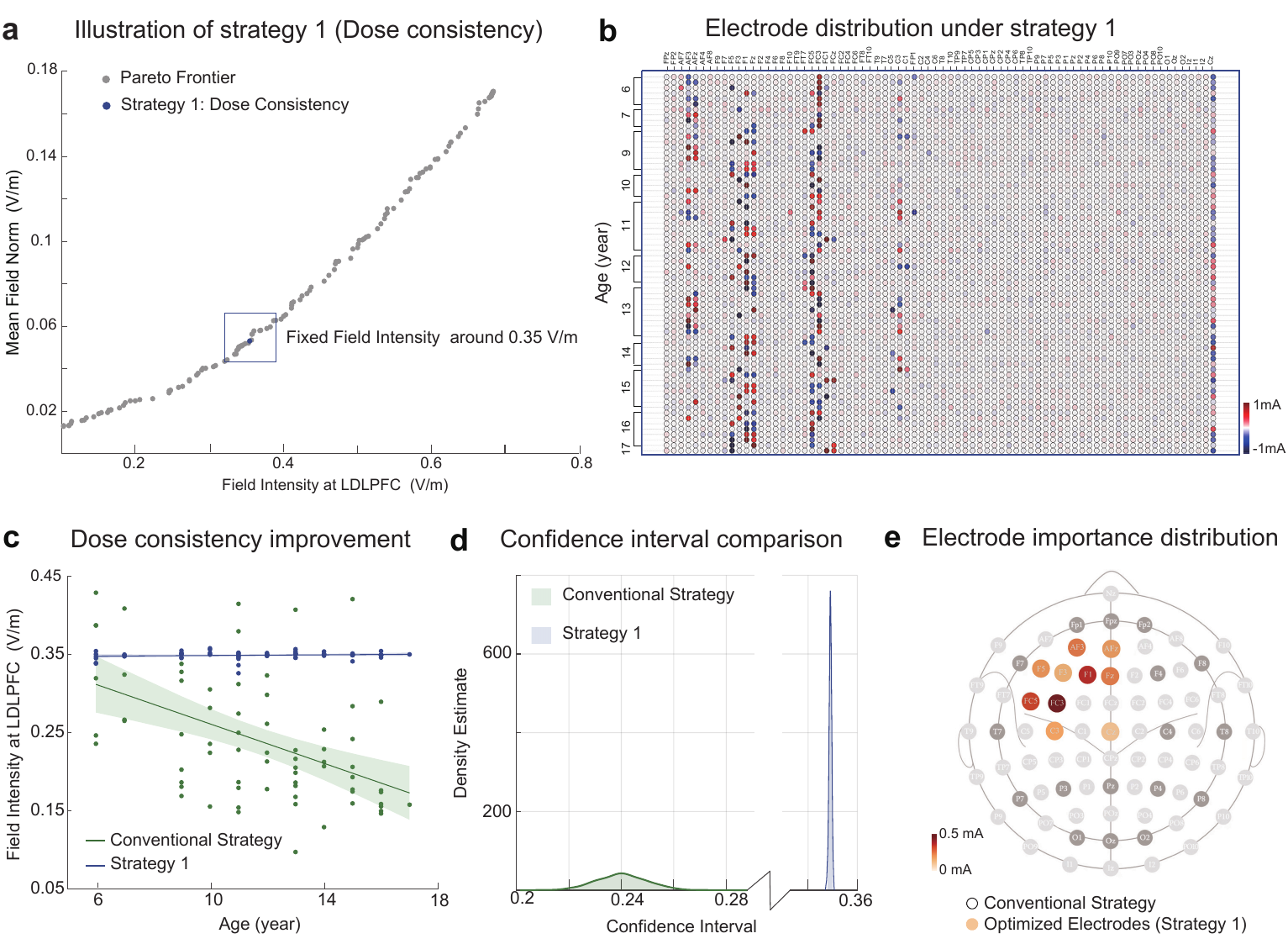}
  \caption{\textbf{LDLPFC-targeted optimization strategies to enhance dose consistency in pediatric HD-tDCS.}
  (a) \textit{Strategy 1 Selection on the Pareto Frontier}: A personalized Pareto frontier for each pediatric subject was derived using the dual-objective optimization framework, balancing field intensity at LDLPFC (x-axis) and mean electric field norm across the brain (y-axis). Strategy 1 (blue) selects a fixed field intensity point ($\sim$0.35 V/m) to stabilize field delivery across individuals and enhance dose consistency.
  (b) \textit{Electrode Current Distributions under Strategy 1}: Heatmaps showing individualized current distributions across 10-10 EEG electrodes under Strategy 1, ordered by subject age. Electrodes are grouped by anatomical regions, with red indicating anodal current and blue indicating cathodal current. Strategy 1 demonstrates age-adaptive flexibility in achieving fixed-intensity targeting.
  (c) \textit{Dose consistency Improvement}:
LDLPFC field intensity (y-axis) vs. age (x-axis) for both conventional montage and Strategy 1. The blue line represents the fixed field intensity at LDLPFC for Strategy 1. The plot demonstrates the ability of Strategy 1 to maintain dose consistency (fixed field intensity) across different ages, with a stability interval (shaded region) around the target intensity.
  (d) \textit{Confidence Interval Comparison (Strategy 1)}:
  KDEs of bootstrap-derived confidence intervals for LDLPFC intensity. The plot shows that Strategy 1 provides more consistent results with narrower confidence intervals, indicating better stability in the field intensity despite variability in electrode configuration.
  (e) \textit{Electrode Importance Distribution under Strategy 1}: 
  Mean absolute current magnitudes aggregated across individuals for Strategy 1. Darker red indicates greater current magnitude. Electrodes used in the traditional montage are outlined in black.
  }
   \label{fig:S3}
\end{figure}

As a supplement to the main-text results on M1-targeted optimization strategies for dose consistency and target engagement (Fig.~\ref{fig:Newfig3}), Fig.~\ref{fig:S3} provides additional details of applying the dual-objective optimization framework targeting LDLPFC in this pediatric cohort for dose consistency and target engagement. Panel (a) shows the Pareto frontier and representative selection of Strategy~1 and Strategy~2. Panel (b) compares electrode current distributions across individuals, highlighting the fixed-intensity control achieved by Strategy~1 and the sparse 4-electrode configuration obtained under Strategy~2 after applying a 0.1 mA threshold. Panel (c–d) demonstrate improved dose consistency under Strategy~1, with stable LDLPFC intensity across age and narrower confidence intervals compared to traditional montages. Panel (e) shows electrode importance distributions under Strategy~1, emphasizing consistent engagement of fronto-central electrodes. Panel (f–g) validate the robustness and efficacy of sparsified Strategy~2, which preserved focality while delivering significantly higher LDLPFC intensities than traditional strategies (***, $p < 0.001$). Finally, panel (h) highlights the electrode importance map of Strategy~2, identifying optimized sites contributing most to efficacy gains.

Together, these results demonstrate how the dual-objective optimization framework produces clinically meaningful strategies for LDLPFC targeting, offering complementary pathways to enhance either dose consistency or target engagement in pediatric HD-tDCS.

\begin{table}[H]
\centering

\caption{Age, sex, and their interaction effects under inverse dual-objective optimization framework targeting Primary Motor Cortex (migraine treatment) under strategy 1 (dose consistency) and strategy 2 (target engagement)}

\begin{adjustbox}{max width=\textwidth}
\begin{tabular}{ccccccccccc}
\toprule
 & $\beta_0$ & $\beta_1$ & $\beta_2$ & $\beta_3$ & $\beta_4$ &  &  & $\Delta$ Deviance & \textit{p}-value  \\
\hline
\rule{0pt}{3pt} \\ 
\textbf{Targeting Motor under Strategy 1} &  &  &  &  &  &  &  &  &   \\
$M_0$     & 30.9566 &  &  &  &  &  &  &  &  \\
$M_1$     & 30.9781 & 30.95659 &  &  &  &  &  & 0 &  0.9986\\
$M_2$     & 30.3436 & 0.0145 &  0.3111 &  &  &  &  &0.0022 & 0.9630 \\
$M_3$     &  30.7509& -0.0203& 0.0323 &  0.0245 &  &  &  &0.0001 & 0.9911 \\
$M_4$    & 30.7509& -0.0203& 0.0323 &  0.0245 &0  &  &  & 0 & 1 \\
\bottomrule
\\
\textbf{Targeting Motor under Strategy 2} &  &  &  &  &  &  &  &  &   \\

$M_0$     & 13.4062 &  &  &  &  &  &  &  &  \\
$M_1$     & 2.7386 & 0.9575 &  &  &  &  &  & 3.2783 &  0.0702\\
$M_2$     & -1.2905 & 1.0504 &  2.0995 &  &  &  &  &3.6523 & 0.0560 \\
$M_3$     &  -1.5882& 1.0788& 2.3136 &  -0.0211 &  &  &  & 0.0004 & 0.9849 \\
$M_4$     & -1.5882& 1.0788& 2.3136 &  -0.0211 &0  &  &  & 0 & 1 \\
\bottomrule
\end{tabular}
\end{adjustbox}

\begin{flushleft}

Statistical analysis based on nested GLMs. The table presents the GLM regression coefficients (\(\beta_0\) to \(\beta_4\)), the difference between the deviance of consecutive nested models ($\Delta$ Deviance), and the corresponding \textit{p}-values ($\chi^2$ test). M0 to M4 represent five different models. (Same conventions as in Table 1.)

\end{flushleft}
\label{tab:glm_personalize_M1}
\end{table}

\begin{table}[H]
\centering

\caption{Age, sex, and their interaction effects under inverse dual-objective optimization framework targeting LDLPFC (cognitive engancement) under strategy 1 (dose consistency) and strategy 2 (target engagement)}

\begin{adjustbox}{max width=\textwidth}
\begin{tabular}{ccccccccccc}
\toprule
 & $\beta_0$ & $\beta_1$ & $\beta_2$ & $\beta_3$ & $\beta_4$ &  &  & $\Delta$ Deviance & \textit{p}-value  \\
\hline
\rule{0pt}{3pt} \\  
\textbf{Targeting LDLPFC under Strategy 1} &  &  &  &  &  &  &  &  &   \\
$M_0$     & 8.1629 &  &  &  &  &  &  &  &  \\
$M_1$     & 8.1589 & 0.00034 &  &  &  &  &  & 0 &  0.9993\\
$M_2$     & 8.1429 & 0.00075 &  0.078 &  &  &  &  &0.00001 & 0.9975 \\
$M_3$    & 8.2051& -0.0046& -0.0346 &  0.0037 &  &  &  &0.00002 & 0.9963 \\
$M_4$     & 8.2051& -0.0046& -0.0346 &  0.0037 & 0  &  &  & 0 & 1 \\
\bottomrule
\\
\textbf{Targeting LDLPFC under Strategy 2} &  &  &  &  &  &  &  &  &   \\

$M_0$    & 1.980 &  &  &  &  &  &  &  &  \\
$M_1$     & 0.167 & 0.160 &  &  &  &  &  & 1.511 &  0.219\\
$M_2$     & -0.449 & 0.173 &  0.346 &  &  &  &  &1.677 & 0.195\\
$M_3$     & -0.415& 0.170 & 0.321 &  0.003 &  &  &  & 0 & 0.993 \\
$M_4$     & -0.415& 0.170 & 0.321 &  0.003 & 0 &  &  & 0 & 0.993 \\
\bottomrule
\end{tabular}
\end{adjustbox}

\begin{flushleft}

Statistical analysis based on nested GLMs. The table presents the GLM regression coefficients (\(\beta_0\) to \(\beta_4\)), the difference between the deviance of consecutive nested models ($\Delta$ Deviance), and the corresponding \textit{p}-values ($\chi^2$ test). M0 to M4 represent five different models. (Same conventions as in Table 1.)

\end{flushleft}
\label{tab:glm_personalize_LDLPFC}
\end{table}

\begin{figure}[htbp]
  \includegraphics[width=0.98\textwidth]{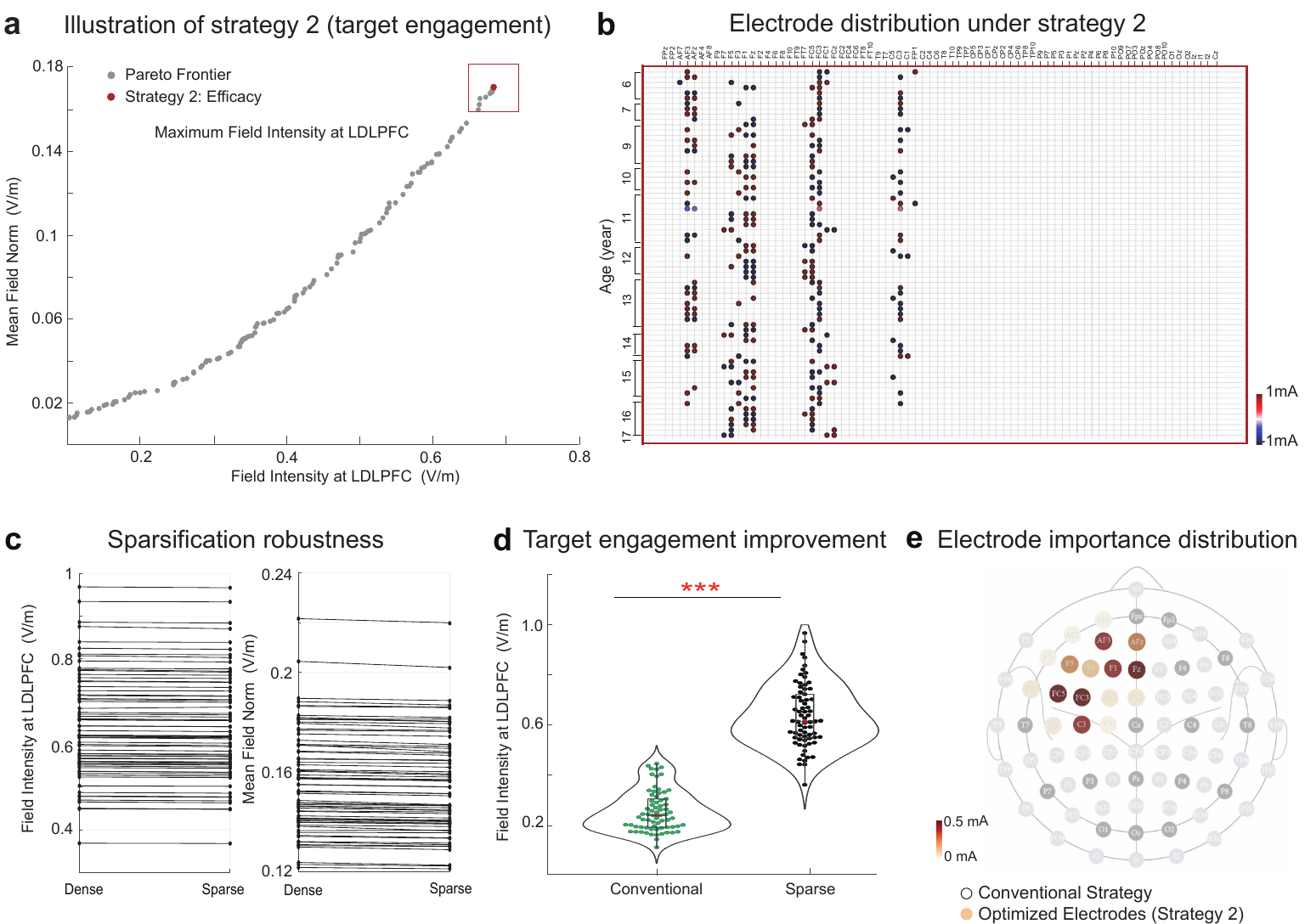}
  \caption{\textbf{LDLPFC-targeted optimization strategies to enhance target engagement in pediatric HD-tDCS.}
  (a) \textit{Strategy 2 Selection on the Pareto Frontier}: A personalized Pareto frontier for each pediatric subject was derived using the dual-objective optimization framework, balancing field intensity at LDLPFC (x-axis) and mean electric field norm across the brain (y-axis). Strategy 2 (red) selects the point with maximal LDLPFC intensity to improve target engagement. Insets highlight representative selections.
  (b) \textit{Electrode Current Distributions under strategy 2}: Heatmaps showing individualized current distributions across 10-10 EEG electrodes under Strategy 2, ordered by subject age. Electrodes are grouped by anatomical regions, with red indicating anodal current and blue indicating cathodal current. Strategy 2 involves a sparse electrode configuration post-processed by thresholding currents below 0.1 mA to increase clinical feasibility.
(c) \textit{Sparsification Robustness}: Dot-and-line plots comparing field intensity at LDLPFC (left) and mean field norm (right) before and after sparsification (Dense vs. Sparse). Dots represent individual data points, while the lines indicate the trends before and after sparsification. No significant differences were observed (n.s.), suggesting that thresholding retains therapeutic field properties while improving clinical feasibility.
  (d) \textit{Target engagement improvement}: Comparison of field intensity at LDLPFC between conventional (green) and Sparse Stimulus Strategy 2 (black) for all subjects. Strategy 2 significantly improves field intensity at LDLPFC (***p < 0.001) compared to conventional strategies, as shown in the violin plot.
  (e) \textit{Electrode Importance Distribution under Strategy 2}: Similar to Fig3S.e, this plot shows the relative importance of electrodes in the optimized electrode layout (Strategy 2) for LDLPFC field intensity. The optimized montage is shown in red, highlighting electrodes that contribute most significantly to increasing LDLPFC intensity.
  }
   \label{fig:S4}
\end{figure}

\end{document}